\documentclass[12pt]{article}

\usepackage{euscript}
\usepackage{amssymb}
\usepackage{amsfonts}
\usepackage{amsbsy}
\usepackage{amsmath}
\usepackage{epsfig}
\usepackage{amsthm}
\usepackage{amscd}
\usepackage{amstext}
\usepackage{graphicx}

\usepackage[pdftex]{hyperref}
\makeatletter
\renewcommand\section{\@startsection {section}{1}{\z@}%
                                   {-3.5ex \@plus -1ex \@minus -.2ex}
                                   {2.3ex \@plus.2ex}%
                                   {\normalfont\large\bfseries}}

\renewcommand\subsection{\@startsection{subsection}{2}{\z@}%
                                     {-3.25ex\@plus -1ex \@minus -.2ex}%
                                     {1.5ex \@plus .2ex}%
                                     {\normalfont\bfseries}}

\newlength{\apb@width}
\newcommand{\autoparbox}[2][c]{\settowidth{\apb@width}{#2}\parbox[#1]{\apb@width}{#2}}

\makeatother

\newcommand{\bea}{\begin{eqnarray}}
\newcommand{\eea}{\end{eqnarray}}
\newcommand{\be}{\begin{equation}}
\newcommand{\ee}{\end{equation}}

\newcommand{\bem}{\begin{pmatrix}}
\newcommand{\eem}{\end{pmatrix}}

\def\a{\alpha}
\def\b{\beta}

\def\e{\epsilon}   
\def\f{\phi}               

\def\im{\mathrm{Im}}
\def\inf{\infty}

\def\l{\lambda}
\def\m{\mu}

\def\o{\omega}  
\def\p{\pi}   
       
\def\t{\tau}
\def\th{\theta}
\def\til{\tilde}

\def\D{\Delta}
\def\F{\Phi}
\def\G{\Gamma}

\def\O{\Omega}

\def\S{\Sigma}

\def\Tr{{\rm Tr}}
\def\U{\Upsilon}

\def \C {{\mathbb C}}

\long\def\symbolfootnote[#1]#2{\begingroup%
\def\thefootnote{\fnsymbol{footnote}}\footnote[#1]{#2}\endgroup}

\begin{document}
\begin{center}
\hfill ITFA-2010-21

\vspace{1cm} 
{\fontsize{19}{0}\bf 
Non-Perturbative Topological Strings \\\vspace{.4cm} And Conformal Blocks}
\vspace{1.5cm}

Miranda C. N. Cheng$~^{\flat, \natural }$, Robbert Dijkgraaf$~^\sharp$ and Cumrun Vafa$^
\diamondsuit \symbolfootnote[1]{On leave from Harvard University.}$

\vspace{0.6cm}

{\it 
$^\flat$Department of Mathematics, Harvard University,
\\
Cambridge, MA 02138, USA \\
$^\natural$Department of Physics, Harvard University,
\\
Cambridge, MA 02138, USA \\
$^\sharp$Institute for Theoretical Physics \& KdV Institute for Mathematics,\\
 University of Amsterdam, The Netherlands \\
$^\diamondsuit$ Center for Theoretical Physics, MIT,
\\
Cambridge, MA 02139, USA \\
}

\vspace{1.0cm}

\end{center}

\begin{abstract}
We give a non-perturbative completion of a class of closed topological string theories in terms of
building blocks of dual open strings. In the specific case where the open string is given by a matrix model these blocks correspond to a choice of integration contour. We then apply this definition to the AGT setup where the dual matrix model has logarithmic potential
and is conjecturally equivalent to Liouville conformal field theory. By studying the natural contours of these matrix integrals and their monodromy properties, we propose a precise map between topological string blocks and Liouville conformal blocks. 
Remarkably, this description makes use of the light-cone diagrams of closed string field theory, where the critical
points of the matrix potential correspond to string interaction points.

\end{abstract}

\pagebreak
\setcounter{page}{1}
\tableofcontents
\section{Introduction}
\label{Introduction}

Topological string amplitudes have played an important role in understanding quantum protected amplitudes in string theory and supersymmetric field theories. More precisely, they capture the F-terms of the corresponding supersymmetric theories. Topological string amplitudes are indexed by the genus of the worldsheet, and amplitudes of different genera compute different physical quantities. For example, the genus zero amplitudes correspond to F-terms in supersymmetric field theories, such as the Yukawa couplings and the superpotential terms. The higher genus amplitudes, on the other hand, involve gravitational corrections. This raises the following question: Does topological string theory have a non-perturbative meaning that goes beyong this term by term identification? And, if so, what is the physical quantity that the full partition function computes?

Various dualities \cite{Gopakumar:1998vy,Gopakumar1999,Dijkgraaf2002} relate the A- and B-model topological string amplitudes to quantities in complex Chern-Simons theories and matrix models respectively. A natural question is hence whether these alternative formulations can help us to find a non-perturbative definition of topological strings. Indeed, both matrix models \cite{David1991,David1993} and Chern-Simons amplitudes 
\cite{Dimofte2009,Witten2010,Witten2010a} do admit non-perturbative definitions, and it is thus natural to ask what the meanings of these non-perturbative completions are in the context of topological string and superstring theories. One feature of such definitions is that, for a given perturbative definition there does not exist a unique non-perturbative completion. Instead there are discrete choices, which we will denote collectively by ${\cal A}$, to be made. For each such choice ${\cal A}$ there is a non-perturbatively defined partition function $Z_{\cal A}^{\text{top}}$ and we are interested in the meaning of these amplitudes $Z_{\cal A}^{\text{top}}$ in the superstring theory. Recalling that topological string amplitudes are interpreted as computing partition functions of superstring theories in specific backgrounds which involve the Taub-NUT spacetime or a 2-dimensional subspace of it, we argue that the choice of ${\cal A}$ translates into the choice of boundary condition for this non-compact spacetime, {\it i.e.} a choice of spacetime branes.

One main motivation for revisiting this question is related to the AGT correspondence \cite{Alday2009} that connects the Nekrasov partition function of four-dimensional ${\cal N}=2$ theories to the conformal blocks  of Liouville (and more generally Toda \cite{Wyllard2009}) theories. These conformal blocks $Z^{\text{CFT}}_{\cal B}$ are labeled by parameters specifying the intermediate channels that we denote collectively by ${\cal B}$. On the other hand, using geometric engineering of the ${\cal N}=2$ theories in the superstring setup, it was shown \cite{Dijkgraaf2009} that these same amplitudes should also correspond to topological string amplitudes represented by matrix model with Penner-like logarithmic potentials. A non-perturbative completion of topological string amplitudes can hence be applied to this special case and should make precise the connection to the conformal blocks of the Liouville theory. In particular, a question to be answered is how the choices ${\cal A}$ of the non-perturbative topological blocks should be mapped to the choices ${\cal B}$ of conformal blocks.

In this paper we argue that the conformal blocks $Z^{\text{CFT}}_{\cal B}$ are in fact a linear combination of the non-perturbative string partition functions $Z_{\cal A}^{\text{top}}$:
$$
Z^{\text{CFT}}_{\cal B}=\sum_{{\cal A}}C_{{\cal B},{\cal A}}\,Z_{\cal A}^{\text{top}}\,,$$
where $C_{{\cal B},{\cal A}}$ are certain constants. By giving such a map, we specify the precise relationship between the Liouville conformal blocks and non-perturbative matrix model amplitudes.

This relation between the conformal blocks of Liouville theory and the periods of the Penner-like matrix integral is one of the interesting by-products of the AGT correspondence. It has been discussed among others in \cite{Schiappa2009,Mironov2010,Mironov2010a,Morozov2010}. Recently it was also considered in the case of genus one surfaces \cite{Maruyoshi:2010pw,Mironov:2010su}. The gauge theory aspect of the relation has been further investigated in, for example, \cite{Eguchi2009,Eguchi2010}. Of course, this problem is directly related to a classical subject in conformal field theory: the contour prescription of the screening charges for minimal models \`a la Dotsenko-Fateev, see {\it e.g.} \cite{Dotsenko1985,Felder1989}. 

As an interesting aspect of defining the topological string blocks $Z_{\cal A}^{\text{top}}$ for Penner-like matrix models, we find that the eigenvalues of the matrix are best represented as points on a light-cone parametrization of a genus zero worldsheet. In this light-cone diagram, the incoming and outgoing tubes correspond to the positions of impurities in the Penner-like potential. The light-cone time $X^+$ is identified with the (real part of the) matrix model potential. In particular, the critical points of the matrix model potential are mapped to the interaction points in the light-cone diagram and vice versa. The basic integration contours of the matrix model are given by straight lines on the light-cone diagram, emanating from the interaction points and going backwards in the light-cone time $X^+$. This connection with light-cone string diagrams is very intriguing and suggests a potentially important connection between light-cone string field theory and ${\cal N}=2$ amplitudes in four dimensions.

The organization of this paper is as follows: In section \ref{Non-Perturbative Topological String Blocks} we formulate a non-perturbative
completion of topological string and its interpretation in terms of superstring amplitudes. This non-perturbative definition can be applied to all topological string theories with a dual open string description. In section \ref{Application} we apply this definition to specific topological string theories that are dual to matrix models with Penner-like logarithmic potential. In section \ref{Topological Strings, Matrix Model, and the AGT Correspondence} we briefly review the content of the AGT correspondence and its relation to the matrix model, and in section \ref{The Liouville Conformal Blocks} we give a dictionary between the parameters in the matrix model and those of Liouville conformal blocks. In this context, in section \ref{The Conformal Blocks versus the Topological String Blocks} we propose a map between the non-perturbative blocks of topological strings and the blocks of the 2d Liouville CFT. In section \ref{Examples} we give examples which illustrate this correspondence. In section \ref{discussions} we conclude with some discussion on future directions. In the Appendix we explicitly demonstrate the relation between degenerate four-point conformal blocks and the corresponding matrix model expressions.

\section{Non-Perturbative Topological String Blocks}
\label{Non-Perturbative Topological String Blocks}
\setcounter{equation}{0}
Topological string amplitudes have an interpretation as computing certain physical amplitudes in superstring theories in the presence of branes. 
The simplest situation where this relation is realized is the following: 
the A-model topological string amplitude computes the partition function of M-theory in the background
$$X\times (S^1\times TN)_q\;,$$
where $X$ is a Calabi-Yau threefold and $(S^1\times TN)_q$ denotes the space obtained by rotating the circle symmetry of the Taub-NUT space by an angle 
$$q=\exp(2\pi ig_s)\;$$ 
as one goes around $S^1$ \cite{Dijkgraaf2007a}.
We can extend this dictionary between the physical and the topological theory to the open topological sectors by introducing A-branes wrapping a Lagrangian subspace $L\!\subset\! X$. In the M-theory setup, adding a topological A-brane corresponds to adding
an M5 brane wrapping the subspace
$$L\times (S^1\times C)_q \subset X\times(S^1\times TN)_q\;,$$
where $C$ is the two-dimensional cigar subspace of $TN$ \cite{DSV-unp,Cecotti2010,Aganagic2009}.
The geometry $(S^1\times C)_q=MC_q$ is the so-called `Melvin Cigar' in \cite{Cecotti2010}.

Reducing on the $S^1$, we can view the M-theory system from the perspective of type IIA superstring theory and obtain a two-dimensional theory on $C$ with ${\cal N}=2$ supersymmetry. The chiral degrees of freedom for this two-dimensional ${\cal N}=2$ theory are associated with the gauge connection $A$ on the Lagrangian submanifold $L$.  Moreover, the superpotential of the two-dimensional theory is simply given by the
Chern-Simons action, up to worldsheet instanton corrections:
$$W(A)/g_s={1\over g_s}\int_L {\rm Tr}(AdA+{2\over 3} A^3)+\text{instanton corrections}\;.$$
In other words, we have
$$Z_{\text {top}}=\int DA\ {\rm exp}(W(A)/g_s)\;.$$
Naively there may appear to be some tension between the following two facts:  in type IIA superstring theory $W$ is the superpotential, whereas here it appears as the action in the topological theory! However, it is known that in order to preserve supersymmetries in two-dimensional ${\cal N}=2$ theories with boundaries, we need to add to the action boundary terms which are exactly given by the integral of the superpotential $W$ evaluated at the boundary \cite{WarnerNucl.Phys.B450:663-6941995}.

In the context of the cigar geometry this has been discussed in \cite{Hori2007}. 
The boundary of the cigar $C$ is a circle, and the path integral localizes on the the field configurations that are constant on it. Hence we arrive at the same formula as above for the physical partition function:
$$Z_{\text{top}}=Z_{\text{IIA}}\;.$$

By mirror symmetry we expect a similar relation between the physical and the topological quantities to hold for the case of the B-model topological string. In this case we have D3-branes wrapping holomorphic curves and filling the cigar subspace of the Taub-NUT geometry.
In particular, we can consider a local Calabi-Yau geometry given by a hypersurface in ${\C}^4$ of the form
$$y^2+W'(x)^2+uv=0$$
with D3-branes wrapped on the holomorphic curves that are described by the conifold-type geometries
localized at critical points of $W(x)$. In this case it
is known \cite{Dijkgraaf2002} that the B-model reduces to a matrix model with potential $W(\F)$:
\be\label{matrix1_IIB} Z_{\text{top}}=Z_{\text{IIB}}=\int D\F\;e^{\Tr \,W(\F)/g_s}\;.\ee
Under this correspondence the rank $N$ of the matrix $\F$ is equal to the total number of branes. 

We will be interested in an expansion near the saddle point in the large $N$ semi-classical limit. A saddle point is specified by a distribution of the eigenvalues among the critical points $p_1, \dotsi, p_n$ of the potential $W$, labeled by the filling fractions $N_1,\dotsi,N_n$ satisfying $\sum N_\ell =N$. This distribution has the interpretation as the number of the branes populating different holomorphic cycles. 

\begin{figure}[h]
  \begin{center}
    \includegraphics[width=5in]{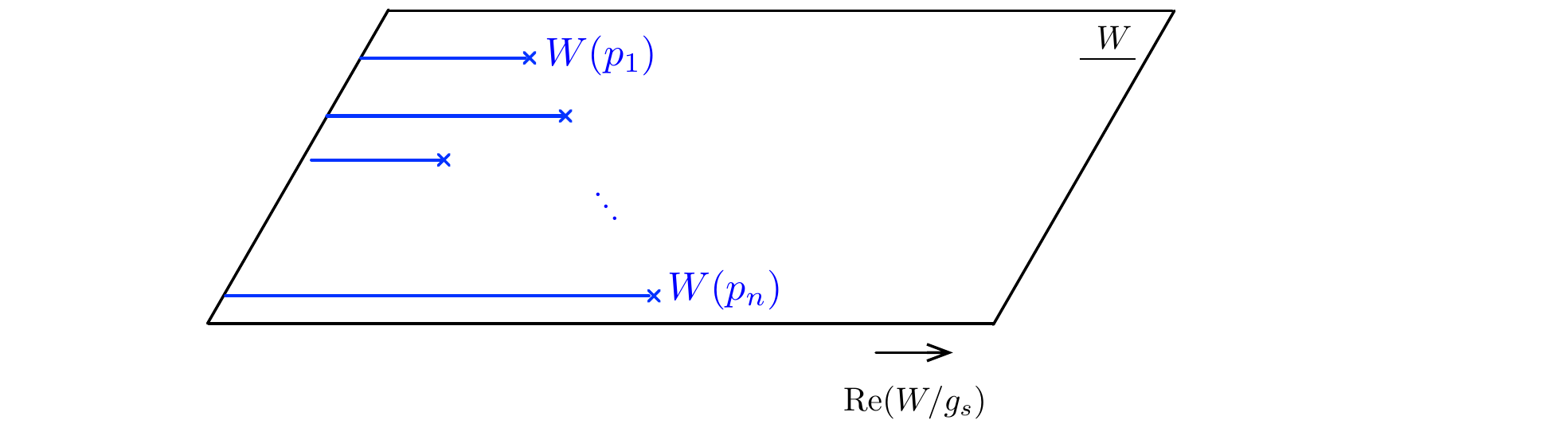} 
  \end{center}
    \caption{\it \footnotesize A natural set of contours $\til C_1,\dotsi, \til C_n$ is given by the pre-images of the straight lines on the $W$-plane. The slope of the straight lines is given by the downward gradient flow of $\text{Re}(W/g_s).$}
    \label{Fig1_Wplane}
\end{figure}

However, there is an intrinsic incompleteness in the above statement. The parameters of the superpotential $W(\F)$ are complex
and the integration over $\F$ is holomorphic.  
Hence, even though the above recipe specifies a perturbative expansion, it does not give an unambiguous non-perturbative answer. 
To remedy this problem, one should specify a contour for the matrix integral. 
In fact, this ambiguity is matched by a corresponding ambiguity in the ${\cal N}=2$ theories in two dimensions. In that context, the ambiguity lies in the choice of a supersymmetric boundary condition. 
As noted in \cite{Hori2007}, to specify a boundary condition for the ${\cal N}=2$ theory in two dimensions, we need to choose a Lagrangian
subspace in the field space.
The allowed Lagrangian submanifolds should satisfy the following conditions. 
First, the projections of the submanifolds in the field space onto the $W$-plane are straight lines emanating from the critical values $W(p_\ell)$. 
Moreover, to ensure the integral is well-defined when the submanifolds have boundaries, we require that
$$|\exp(W/g_s)|\rightarrow 0$$
at these boundaries. These two conditions are illustrated in Figure \ref{Fig1_Wplane}.

A way to visualize these straight lines on the original Riemann surface is via the following quasi-conformal mapping  
$$
ds^2 = |dw|^2 = |\f_{zz}|\,|dz|^2 \;,
$$
associated with a so-called Jenkins-Strebel holomorphic quadratic differential $\f_{zz}(z)(dz)^2$ given in terms of the matrix potential as
$$
 \f_{zz} (dz)^2 = (dw)^2 = \big(\frac{1}{g_s} dW\big)^2\;.
$$
Clearly, such a quasi-conformal mapping defines a metric that is flat everywhere on the Riemann surface except at the zeros and poles of $\f_{zz}$, where the curvature  is singular. By definition, the zeros are exactly the critical points of the matrix potential $W(z)$ and their images under the $W$-map are the points where the Lagrangian submanifolds emanate from. 


The choice of the Lagrangian submanifold makes precise the way the complex integral over the chiral fields should be performed, and as a result the answer for the partition function depends on this choice. In other words, on both the topological and physical sides, a choice of a Lagrangian submanifold is required to define the amplitude unambiguously. In the first case, the Lagrangian submanifold plays the role of the integration contour, while in the latter case it completes the definition of the theory by providing a supersymmetric boundary condition.  

Let us summarize this main point that is crucial to the rest of the paper: {\it Both the topological and physical amplitudes are uniquely defined once a choice of the Lagrangian submanifold is made. It is hence natural to identify this choice of the Lagrangian submanifold as a choice of a non-perturbative completion of the topological string amplitudes.}

In other words, for a given choice of the topological string coupling constant $g_s$ and the parameters of the superpotential $W$, the data of brane distribution $N_\ell$ that characterizes the matrix model saddle points also leads to a non-perturbative answer for the topological amplitude 
$$Z(N_1,\ldots,N_n) = \text{non-perturbative topological block labeled by }\{N_1,\ldots, N_n\}\;.
$$ 
The fact that the matrix model integrals with complex potentials can be defined using such complex contours has been known for a long time \cite{David1991,David1993}. More recently they were discussed in the context of a non-perturbative formulation of topological strings in \cite{Marino2007a,Marino2008,Eynard2008}. In the context of complex Chern-Simons, the idea of associating a contour to a given classical solution has appeared in \cite{Dimofte2009}. 
Moreover, similar Lagrangian contours, which are infinite dimensional in this context, also feature in the discussion on the complex Chern-Simons path integral \cite{Witten2010}.

However, such a `topological string block' $Z(N_1,\ldots,N_n)$ is not a continuous function of the parameters of the superpotential. In particular, as can be seen in Figure \ref{Fig1_Wplane}, it jumps at special values of the parameters such that there are two critical points $p_{\ell_1},p_{\ell_2}$ with $$\im\left(\frac{W(p_{\ell_1})-W(p_{\ell_2})}{g_s}\right)=0\;.$$ 
These jumps in the corresponding cycles result in the Stokes' phenomenon for the integral described by the Picard-Lefschetz theory, and cause non-trivial monodromy transformations when the parameters move around in the complex plane. This property is of course reminiscent of that of conformal blocks in two-dimensional conformal theory. In this paper we will demonstrate how this analogy, when suitably interpreted, can be made into an equality in the context of the Liouville conformal theory interpreted as topological string theory \cite{Dijkgraaf2009}.

\section{Application To Penner-Type Matrix Models}
\label{Application}
\setcounter{equation}{0}

In the previous section we have given a definition of non-perturbative topological string blocks through matrix integrals. In this section we would like to apply such a definition to Penner-like matrix models with logarithmic potentials. As we will review later, this type of matrix models is of special interest due to its relation to the Liouville conformal blocks. Hence an understanding of the corresponding topological string blocks will be a prerequisite for a concrete realization of \cite{Dijkgraaf2009}, which connects certain topological string amplitudes to Liouville conformal blocks in the context of the AGT correspondence \cite{Alday2009}. To keep the discussion explicit we will focus on the genus zero conformal blocks of the $SL(2)$ Liouville theory. Similar consideration should also apply to the more general situations of higher genus surfaces and higher rank gauge groups discussed in \cite{Dijkgraaf2009}.

 We are interested in the matrix models with the following type of logarithmic potential 
 \be\label{genus_0_potential}
{W(z)}/{g_s}=\sum_{\ell=1}^{n+1}{m_\ell}\, \log(z-z_\ell)  \;, 
\ee
corresponding to the matrix integral
\be\label{matrix_integral_1}
Z =\idotsint \prod_{i=1}^N du_i \prod_{1\leq i < j \leq N} (u_j-u_i)^{2} \,\prod_{i=1}^N \prod_{\ell=1}^{n+1} (u_i-z_\ell)^{m_\ell}\;.
\ee
Here we have written the integral in terms of the eigenvalues $u_1,\dotsi, u_N$ of the matrix $\F_{N\times N}$.

As discussed in the previous section, we are interested in the gradient flows of Re$(W/g_s)$, which are by definition straight lines on the Riemann surface with respect to the metric 
$$
ds^2 = \left\vert\frac{1}{g_s} dW\right\vert^2\;.
$$
For the type of matrix model under consideration, the quasi-conformal transformation is particularly nice and in fact gives a parametrization of the Riemann surface as a scattering amplitude of closed string field theory in the light-cone gauge \cite{Giddings1986}. To see this, let us first study the zeros and poles of the quadratic differential $\f_{zz}dz^2=(dW/g_s)^2$, or more precisely by the one-form $\omega$ given by $\f=\omega^2$. 

Recall that a meromorphic one-form on a genus $g$ Riemann surface has $2g-2$ `net' number of zeros (the number of zeros minus the number of poles). There is a particularly relevant one-form $\omega$, introduced in \cite{Giddings1987}, that is uniquely determined by the following data: 

(1) the (real) residues $m_\ell$ at the poles $z_\ell$
$$
m_\ell = \oint_{z_\ell} \omega,
$$
which should necessarily satisfy $\sum_\ell m_\ell = 0$, together with 

(2) the condition that the periods
$$
\oint_C \omega
$$
over the one-cycles $C$ on the surface are all purely imaginary.

 Given such a one-form $\omega$, one can define a local coordinate $w$ by
$$
w = \int^z \omega.
$$
This light-cone coordinate $w$ is related to the matrix model potential as
$$
w = {1\over g_s} W(z).
$$

In the light-cone string diagram the poles of $\omega$ describe the incoming and outgoing external strings. If we have
$$
\omega(z) \sim {m_\ell \over z - z_\ell},
$$
then correspondingly
$$
w(z) \sim m_\ell \log(z-z_\ell) \to \pm \infty,
$$
depending on the sign of $m_\ell$. 

Similarly, the zeroes $p_\ell$ of $\omega$ correspond to an interaction point where two incoming closed strings merge into an outgoing one, or one incoming string splits into two outgoing ones. Let us recall why this is the case. If
$$
\omega(z) \sim (z-p_\ell)dz\,,
$$
this implies that
$$
w(z) \sim (z-p_\ell)^2.
$$
Clearly if $z$ encircles the zero $p_\ell$ once, the coordinate $w$ is rotated over $4\pi$. So we have a surplus angle that comes with the conical singularity of a pair of pants geometry.

In the present case of genus zero, the poles of the one-form $\omega$ are the $n+2$ poles located at the finite values $z=z_{1, \ldots,n+1}$ together with an extra pole at $z_{n+2}=\inf$. The zeroes of $\omega$ are located at the $n$ critical points $p_{1, \ldots,n}$ of the matrix potential satisfying
$$
  \frac{dW}{dz}\Big\lvert_{z=p_\ell} =  \sum_{k=1}^{n+1}  \frac{m_{k}}{p_{\ell}-z_{k}} =0\;.
$$

\begin{figure}[h]
  \begin{center}
    \includegraphics[width=5.5in]{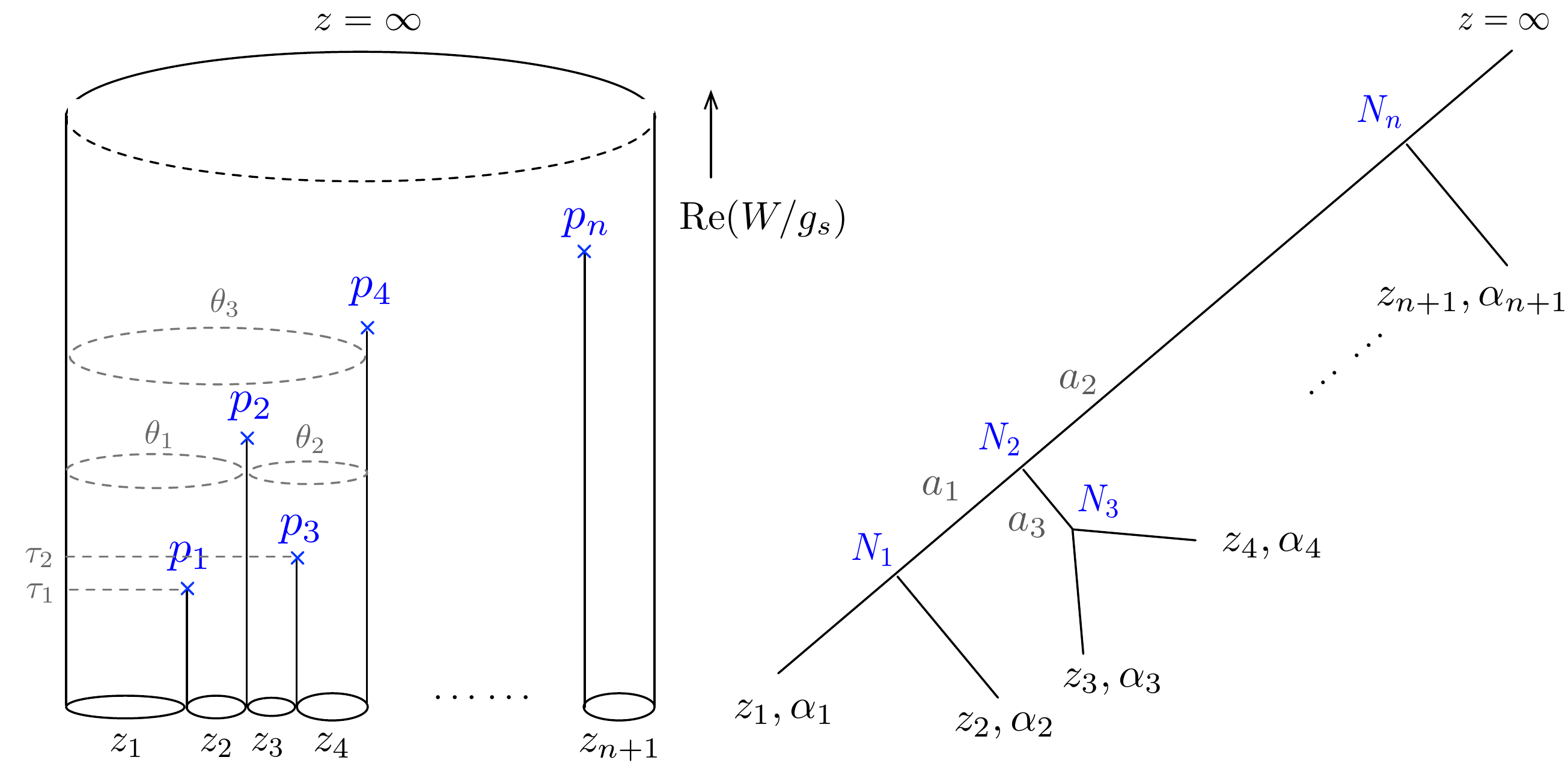} 
  \end{center}
    \caption{\it \footnotesize An example of a light-cone diagram for a given matrix potential $W$. The locations $z_\ell$ of the poles of $dW$ determine the twist angles $\th_\ell$ and the interaction times $\tau_\ell$, while the residues $m_\ell$ determine the size of the closed strings. 
    Such a light-cone diagram gives a pants decomposition of the genus zero Riemann surface and thereby determines a tree structure. Together with the data of the distribution $\{N_1,\ldots,N_n\}$ of the matrix eigenvalues among the critical points $\{p_1,\ldots,p_n\}$ of the matrix potential, this tree specifies a corresponding conformal block. }
    \label{many_legs}
\end{figure}

The arrow of time of the light-cone is given by increasing Re$(W/g_s)$, as shown in Figure \ref{many_legs}. Such a genus zero light-cone diagram is specified by the following data: the external momenta $\{m_1,\ldots,m_{n+1}\}$, the times $\{\t_1,\ldots,\t_{n}\}$ of interaction, and the twisting angles $\{\th_1,\ldots,\th_{n-1}\}$ with which an intermediate closed string is twisted before rejoining with the rest of the diagram \cite{Giddings1987}. Fixing the initial time to be $\t_1=0$, we see that the $\t$'s and $\th$'s together give the $(n-1)$ complex variables corresponding to the locations of the poles $z_1,\ldots,z_{n+1},z_{n+2}=\inf$ up to $SL(2,\C)$ equivalence.


In order to obtain an unambiguous and uniform description of the contour for the matrix model integral (\ref{matrix_integral_1}), we will consider the potential (\ref{genus_0_potential}) with all momenta $m_{\ell}$ being real and positive. The answer for the integral can then be analytically continued to other ranges of the parameters. For such cases, the quasi-conformal map described above leads to light-cone diagrams of the kind depicted in Figure \ref{many_legs}. In other words, we consider scattering diagrams with $n+1$ incoming and one outgoing closed string. As a result, in our matrix model discussion, the pole of $dW$ at the infinity of the genus zero curve is in some sense more special than all the other poles at $z_1,\ldots,z_{n+1}$.

As mentioned in section \ref{Non-Perturbative Topological String Blocks}, we would like to associate a saddle point, labeled by the eigenvalue distribution 
$ \{N_1,\ldots,N_{n}\}$, a matrix model integral (\ref{matrix_integral_1}) with a specific contour. 
In the language of the light-cone diagram (see Figure \ref{many_legs}), this means we should associate a class of contour to each interaction point of the strings. As discussed in section \ref{Non-Perturbative Topological String Blocks}, there is a very natural way to specify these contours from the matrix model point of view: they are given by the gradient flow of Re$(W/g_s)$. In the present case they are simply given by the downward-flowing (backward in time) straight lines emanating from the interaction points on the light-cone diagram. In particular,  the corresponding contours are not closed cycles.

Concretely, for a given potential $W$ of the form (\ref{genus_0_potential}) and thereby a given light-cone diagram, label the poles $z_1,\ldots,z_{n+1}$ and zeros $p_1,\ldots,p_n$ of the corresponding quadratic differential in the order depicted in Figure \ref{many_legs}. Then the contour associated in the abovementioned way to the $\ell$-th critical point is the line segment going between the point $z_{\ell}$ and $z_{\ell+1}$ and passing through the critical point $p_\ell$. We choose the orientation and denote by $\til C_\ell$ the line segment going from $z_{\ell+1}$ to $z_{\ell}$. As a result, to a given distribution of eigenvalues $N_\ell$ among the critical points $p_\ell$, we can associate a matrix integral (\ref{matrix_integral_1}) with the following contour 
\be\label{basis1}
\otimes_{\ell=1}^n \til C_\ell^{\otimes N_\ell}  \;\quad,\quad\til C_ \ell = [z_{\ell+1},z_{\ell}]\;.
\ee

As we will discuss in more details later, when the momenta $m$'s are not all integers and when we consider the $\b$-ensembles generalization of the matrix model, the genus zero curve on which the matrix eigenvalues live is in fact multi-sheeted. The matrix integrals with the above prescription of contours transform non-trivially when changing from sheet to sheet and in fact mix with one another under such monodromy actions. One might hence deem it natural to use an alternative basis for matrix integrals on which the monodromy transformations act diagonally. In fact this property is mandatory from the point of view of the conformal blocks. We will construct such an alternative basis and relate the corresponding matrix integrals to conformal blocks in section \ref{The Conformal Blocks versus the Topological String Blocks}.

\section{Logarithmic Matrix Models and the AGT Correspondence}
\label{Topological Strings, Matrix Model, and the AGT Correspondence}
\setcounter{equation}{0}
In this section we will briefly review the relation between certain 2d conformal blocks and the Nekrasov partition functions for certain ${\cal N}=2$, $d=4$ superconformal field theories \cite{Alday2009}, and how this relation can be understood through 
their connections to topological string theory and matrix model \cite{Dijkgraaf2009}. 

In \cite{Alday2009}, the authors consider a class of ${\cal N}\!=\!2$, $d\!=\!4$ superconformal field theories, which can be obtained from compactifying the six-dimensional M5 brane ${\cal N}=(2,0)$ superconformal field theories on punctured Riemann surfaces $\S$ \cite{Gaiotto2009}. These theories admit an A-D-E classification. These authors propose that, for the case of $SU(2)$ theories corresponding to two M5 branes, the Nekrasov partition functions in the $\O$-background are essentially given by certain conformal blocks of the Liouville conformal field theory. Underlying both objects is the genus $g$ Riemann surface $\S$ with $n+2$ punctures, and its pants decomposition determines a weakly coupled Lagrangian description on the gauge theory side and a specific channel on the 2d CFT side. More specifically, the sewing parameter of a neck connecting two pairs of pants is related to the gauge coupling constant of the corresponding $SU(2)$ gauge group. Moreover, the external momenta of the Liouville conformal block correspond to the hypermultiplets masses of the gauge theory, while the internal momenta map to the Coulomb parameters. Finally, the two real numbers $\e_1,\e_2$ characterizing the $\O$-background fix the parameter $b$ of the Liouville theory as
$$
b^2 = \e_1/\e_2 \;. 
$$
Later this proposal was generalized to the $A_r$ 4d SCFT/2d Toda theory for $r>1$ in \cite{Wyllard2009}. 

An explanation for this correspondence was suggested in \cite{Dijkgraaf2009}. In this way of understanding the AGT correspondence, the matrix model plays the role of a bridge between the 2d and the 4d theories. On the one hand, it has long been known that the collective dynamics of the eigenvalues of a Hermitian matrix model can be described by a chiral conformal field theory with a Liouville-type interaction \cite{Marshakov1991,Kostov2007}. On the other hand, the (refined) topological string partition function on the local Calabi-Yau geometry which engineers the gauge theory is known (after $\beta$-deformation) to compute the Nekrasov partition function, and at the same time also has a description in terms of matrix model amplitude via the open/closed geometric transition \cite{Dijkgraaf2002}.  

More concretely, in the genus zero case, from the brane insertion picture it was clear that a logarithmic matrix model with potential of the form (\ref{genus_0_potential}) should be the relevant one for connecting the 2d and 4d quantities, and the parameters $m_\ell$'s should be related to the hypermultiplet mass parameters on the 4d SCFT side, or the external momenta on the 2d conformal block side. Moreover, to describe a general $\O$-background, corresponding to a refined topological string theory, we should consider the $\b$-ensemble deformation of the matrix model, turning the measure in the matrix integral from (\ref{matrix_integral_1}) into 
\be\label{matrix_integral_beta}
Z =\idotsint \prod_{i=1}^N du_i \prod_{1\leq i < j \leq N} (u_j-u_i)^{2\b} \,\prod_{i=1}^N \prod_{\ell=1}^{n+1} (u_i-z_\ell)^{m_\ell}\;,
\ee
with $$\b =- \e_1/\e_2\quad,\quad g_s^2 =-\e_1\e_2\;.$$
At the same time, in the context of relating matrix model to a Liouville-like chiral CFT, the eigenvalue integral plays the role of the a screening operator. Hence, the data of eigenvalue distribution $\{N_1,\ldots,N_n\}$ should be related to the Coulomb parameters on the 4d side, or the internal momenta on the 2d side. Indeed, the $m_{\ell=1,\ldots,n+1}$ parameters in the matrix potential together with the numbers $N_{\ell=1,\ldots,n}$ of eigenvalues clustered around a given critical point make together the $2n+1$ parameters specifying the $n+2$ external and $n-1$ internal momenta in a conformal block. 

After defining the non-perturbative topological string blocks for general topological string theories with a matrix model description, the second goal of this paper is to use this non-perturbative definition to make concrete the map between such blocks and the Liouville conformal blocks in the context reviewed above. For this purpose, we first have to give a precise map between the matrix model parameters and those of the conformal blocks. 
This will be the topic of the next section. Second we have to give the map between the topological blocks we defined earlier and the conformal blocks with the corresponding parameters, which will be the topic of section \ref{The Conformal Blocks versus the Topological String Blocks}.

\section{From Matrix Models to Conformal Blocks}
\label{The Liouville Conformal Blocks}
\setcounter{equation}{0}

In this section we will see how to associate a genus zero Liouville conformal block to a matrix model amplitude. The former is labeled by a tree structure, the momenta $\a_{1},\ldots,\a_{n+2}$ and the locations $z_1,\ldots,z_{n+2}$ of the operator insertions $V_{\a_\ell}(z_\ell)$, and the internal momenta $a_{1},\ldots,a_{n-1}$. The latter is given by the parameters  $m_{1},\ldots,m_{n+1}$ and $z_{1},\ldots,z_{n+1}$ specifying the potential (\ref{genus_0_potential}), the rank of the matrix $N$, and the $n-1$ independent filling numbers $\{N_1,\ldots,N_n\}$ labeling the saddle-point of the matrix model path integral. 

Also for this purpose we find the light-cone diagram introduced in section \ref{Application} illuminating: For a given matrix potential $W(z)$, the corresponding quasi-conformal map gives a unique pants decomposition of the genus zero curve. As shown in Figure \ref{many_legs}, the skeleton of such a pants decomposition gives rise to a natural tree structure. We would like to identify this tree structure with that of the conformal block. Moreover, as suggested by the picture, the locations of the poles of the quadratic differential can naturally be identified with the locations of the operator insertions in the Liouville block. In particular, we identify the position of the last operator insertion with the pole of $dW$ at infinity and write $z_{n+2}=\inf$. By the same token, the parameters $m_{1,\ldots,n+1}$ and the first $n+1$ external momenta $\a_{1,\ldots,n+1}$ should have a simple linear relation. 
More precisely, comparing the matrix integral (\ref{matrix_integral_beta}) with the holomorphic part of the free boson correlators 
$$
\langle V_{\a_1}(z_1,\bar z_1) \dotsi V_{\a_{n+2}}(z_{n+2},\bar z_{n+2}) \rangle =\prod_{1\leq k<\ell\leq n+2} \lvert z_{k\ell}\rvert^{-4\a_k \a_\ell}\;,\; z_{k\ell} =z_\ell-z_k
$$ 
in the context of the Coulomb gas treatment of the Liouville conformal theory, we get the following dictionary between the matrix model parameters on the one side and the CFT data on the other: 
$$
\b=-b^2\quad,\quad m_{\ell} = -2 b \a_{\ell}\;,\; \ell =1,\ldots,n+1\;,
$$
where $b$ is the parameter in the Liouville action. The corresponding background charge and central charge of the Liouville theory are given by
$$
Q= b + b^{-1}\quad,\quad c=1+ 6Q^2\;.
$$

Furthermore, a comparison of the matrix integral (\ref{matrix_integral_beta}) with the free boson correlators shows there is an overall factor difference
$$
\text{conformal block} = \prod_{1\leq k<\ell\leq n+1} (z_\ell-z_k)^{m_k m_\ell/2\b} \times \text{matrix integral}
$$
between the conformal block and the corresponding matrix integral. Notice that this overall factor is independent of the internal momenta of the conformal block, or equivalently the eigenvalue distribution of the matrix model. Therefore the presence of this extra factor does not influence our discussion on the relation between of the conformal and topological string blocks, since in such a discussion we always fix the external momenta, or equivalently the integrand of the matrix integral on the matrix model side. 

\begin{figure}[h]
  \begin{center}
    \includegraphics[width=4in]{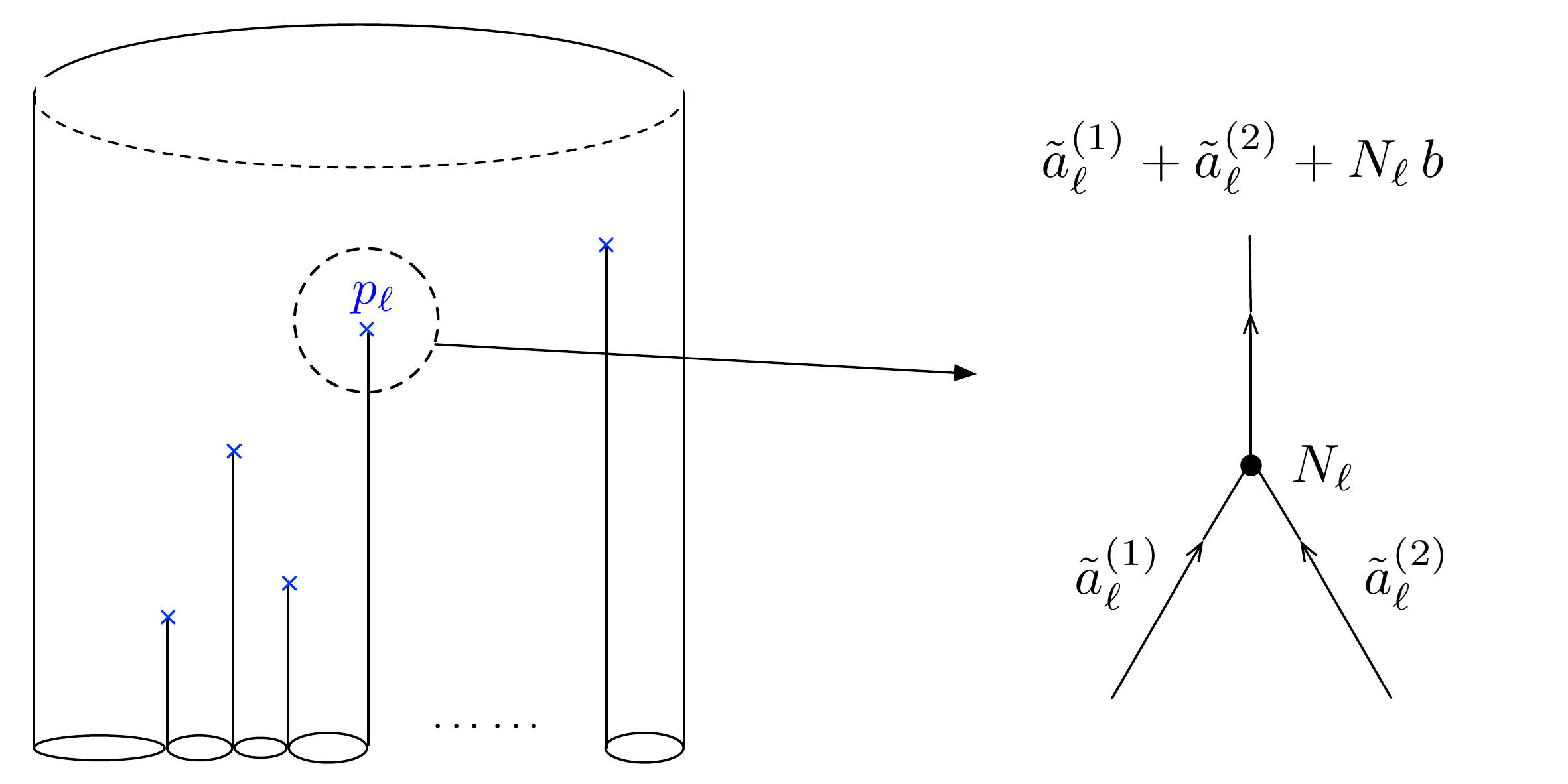} 
  \end{center}
    \caption{\it \footnotesize The momentum violation at each vertex is given by the number of eigenvalues at the corresponding critical point, directed towards increasing light-cone time. }
    \label{momenta_violation}
\end{figure}

Finally we would like to map the data of matrix model saddle-point $\{N_{1,\ldots,n}\}$ to the parameters on the conformal block side. As mentioned earlier, in the context of the chiral Liouville theory, a matrix eigenvalue integral corresponds to adding a screening charge, changing total momenta by an amount $\D \a = b$. At the same time, according to our map between the closed string scattering diagram and the conformal block tree, a matrix critical point where an eigenvalue can be deposited is mapped to a vertex of the conformal block tree. Hence it is natural to identify the number of eigenvalues $N_{\ell}$ at a given critical point $p_\ell$ with the amount of momentum non-conservation at the corresponding vertex in the conformal block tree, as illustrated in Figure \ref{momenta_violation}. 

As a result, the $n-1$ internal momenta $a_{\ell=1,2,\ldots,n-1}$ are determined by the following momenta violation condition at each vertex vertices of the conformal block
\be
\til \a_\ell^{(1)} +\til  \a_\ell ^{(2)} +\til  \a_\ell ^{(3)} = Q - N_\ell \,b\;.
\ee
Here $\til \a_\ell ^{(1,2,3)}$ denote the three momenta coming into the $\ell$-th vertex of the conformal block. Equivelently, when choosing $\til \a_\ell^{(1,2)}$ to be incoming and $\til\a_\ell^{(3) \ast} $ to be outgoing at the $\ell$-th vertex, we have 
$$
\til\a_\ell^{(3) \ast} =\til \a_\ell^{(1)}+\til \a_\ell^{(2)} + N_\ell \,b\;
$$
as shown in Figure \ref{momenta_violation}.
Recall that flipping the orientation of a momentum $\a$ in the presence of the background charge amounts to taking $$\a^\ast = Q-\a\;,$$ which leaves the conformal dimension
$$
\D_\a = \D_{\a^\ast} = \a (Q-\a)
$$
of the corresponding vertex operator $V_\a$ invariant. For example, when the direction of all momenta  in Figure \ref{many_legs} is chosen to flow forward in time to the vertex $z=\inf$, the first three internal momenta are
$$
a_1=\a_1+\a_2+N_1 b \;,\; a_3=\a_3+\a_4+N_3 b\;,\; a_2 = a_1+a_3+N_2 b\;.
$$

Notice that this rule for the momenta violation at each vertex also implies that the last external momentum $\a_{n+2}$ of the operator inserted at $z_{n+2}=\inf$ is given by $m_{1,\ldots,n+1}$ and the rank of the matrix as
$$
\sum_{\ell=1}^{n+2} \a_\ell = Q- Nb\;.
$$
This completes the map between the parameters on the matrix model and on the conformal block sides.

\section{Topological String Blocks Versus Conformal Blocks}
\label{The Conformal Blocks versus the Topological String Blocks}
\setcounter{equation}{0}
In this section, by studying the monodromy transformations of different matrix model blocks, we will determine a family of contours appropriate for the purpose of computing CFT conformal blocks. See also  \cite{Dotsenko1985} and \cite{Felder1989} for some earlier discussions on the integral representation of conformal blocks in the context of minimal models.

As alluded to earlier, the different ways of gluing the pairs of pants introduce extra subtleties when defining the contours. 
When the parameters $m_\ell$ and $\b$ are not integral, the contour $\til C_\ell=[z_{\ell+1},z_{\ell}]$ which cuts through a gluing curve along which different pants are glued together will not be invariant under the corresponding Dehn twist operation. This is because the presence of the branch cuts has rendered the genus zero Riemann surface multi-sheeted. As a result, the basis (\ref{basis1}) for the matrix integral with a given potential is in general not a diagonal basis under the monodromy transformation of the various parameters in the matrix model potential, in particular the locations  $z_\ell$ of the poles. 

Therefore, in order to unambiguously specify a contour for the matrix integral with a given potential $W$ and eigenvalue distribution $\{N_1,\ldots,N_n\}$, it seems natural to look for an alternative basis on which the monodromy operations act diagonally. More concretely, we would like to combine the contour $\til C_\ell$ and its images under the monodromy operations, in such a way that the combined contour $C_\ell$ furnishes such a diagonal basis. We will see that there exists a unique (up to overall multiplicative factors) contour $C_\ell$ satisfying this condition. 

\begin{figure}[h]
  \begin{center}
    \includegraphics[width=5.5in]{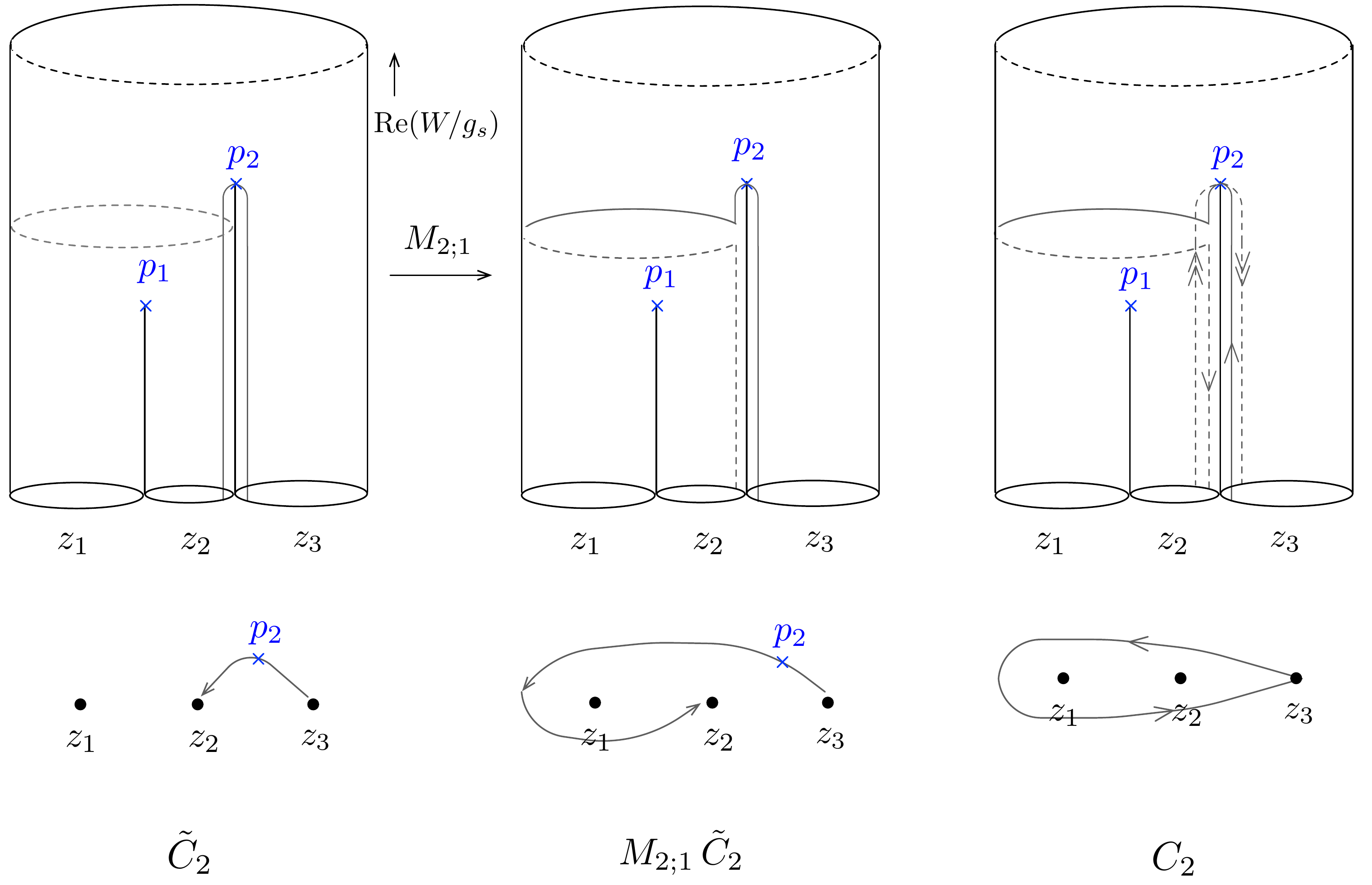} 
  \end{center}
    \caption{\it \footnotesize(a) The contour $\til C_2$ given by the pre-images of straight line in the $W$ plane. (b) Its image under monodromy transformation $M_{2;1}$ that takes the location $z_2$ of the pole around $z_1$ in a counterclockwise orientation. (c) The combination $C_2$ of the two contours that is invariant under the monodromy transformation. }
    \label{monodromy_4pnt_ex}
\end{figure}

We are now ready to analyze the action of the Dehn twist operation on our contours $\til C_{\ell}$ defined in section \ref{Non-Perturbative Topological String Blocks} and \ref{Application}. 
As is suggested by the picture (see, for example, Figure \ref{monodromy_4pnt_ex}), the contour $\til C_\ell$ can be seen as the union of two line segments connecting the critical point $p_\ell$ to the poles $z_\ell$ and $z_{\ell+1}$ respectively. 
Depending on the light-cone diagram, none, one, or both of these two line segments will cut through some gluing curves. In the first case, there is no Dehn twist to be performed and the contour $\til C_\ell$ we defined earlier is also the unambiguous contour $$C_\ell=\til C_\ell$$ we are looking for. See Figure \ref{3point} for such an example.

To discuss the other cases, let us first look at the simple example shown in Figure \ref{monodromy_4pnt_ex}. The contour $\til C_2$ cuts through the gluing curve along which the tubes extending from $z_{1}, \,z_{2}$ are glued to the rest of the light-cone diagram. Hence, as can be seen from the Figure \ref{monodromy_4pnt_ex}, our original contour $\til C_2$ is not invariant under the monodromy $M_{2;1}$ that takes the point $z_{2}$ around the point $z_{1}$ in the counter-clockwise orientation before returning to the original position. To find the invariant contour, let us first observe that the image of $\til C_2$ under this monodromy is 
$$ M_{2;1}\til C_2 = \til C_2 + L\;,$$
where  $L=(e^{\f_2}-e^{\f_1+\f_2})\til C_1$ is a loop going through $z_2$ and circling around $z_1$, and $e^{\f_\ell}$ is the phase incurred by passing counter-clockwise through the brach cut connecting the point $z_\ell$ and infinity. More precisely, in the basis given by the contours $\{\til C_1, \til C_2\}$, the monodromy matrix reads
$$
M_{2;1}\bem \til C_1 \\ \til C_2\eem  
= \bem  e^{\f_1+\f_2}&0 \\  e^{\f_2}-e^{\f_1+\f_2}&1 \eem \;\bem\til C_1 \\ \til C_2\eem\;.
$$ 
Hence we have 
$$
(M_{2;1})^k \til C_2  = \til C_2 + (1+q+\ldots+q^{k-1}) L \;,\; q=e^{\f_1+\f_2}. 
$$ 
We would like to take a combination $\sum_k c_k (M_{2;1})^k  \til C_2$ of the images of the contour $\til C_2$ that is invariant under the monodromy. It is not hard to see that the contour 
$$C_2=(M_{2;1}-q)\til C_2\;,$$
which can be combined into a loop passing through the point $z_3$ and circling the points $z_1,z_2$, 
is the unique invariant combination up to a multiplicative factor. Hence we arrive at the contour $C_2$ shown in Figure \ref{monodromy_4pnt_ex} which has the desired property under monodromy transformation. 

\begin{figure}[h]
  \begin{center}
    \includegraphics[width=5.5in]{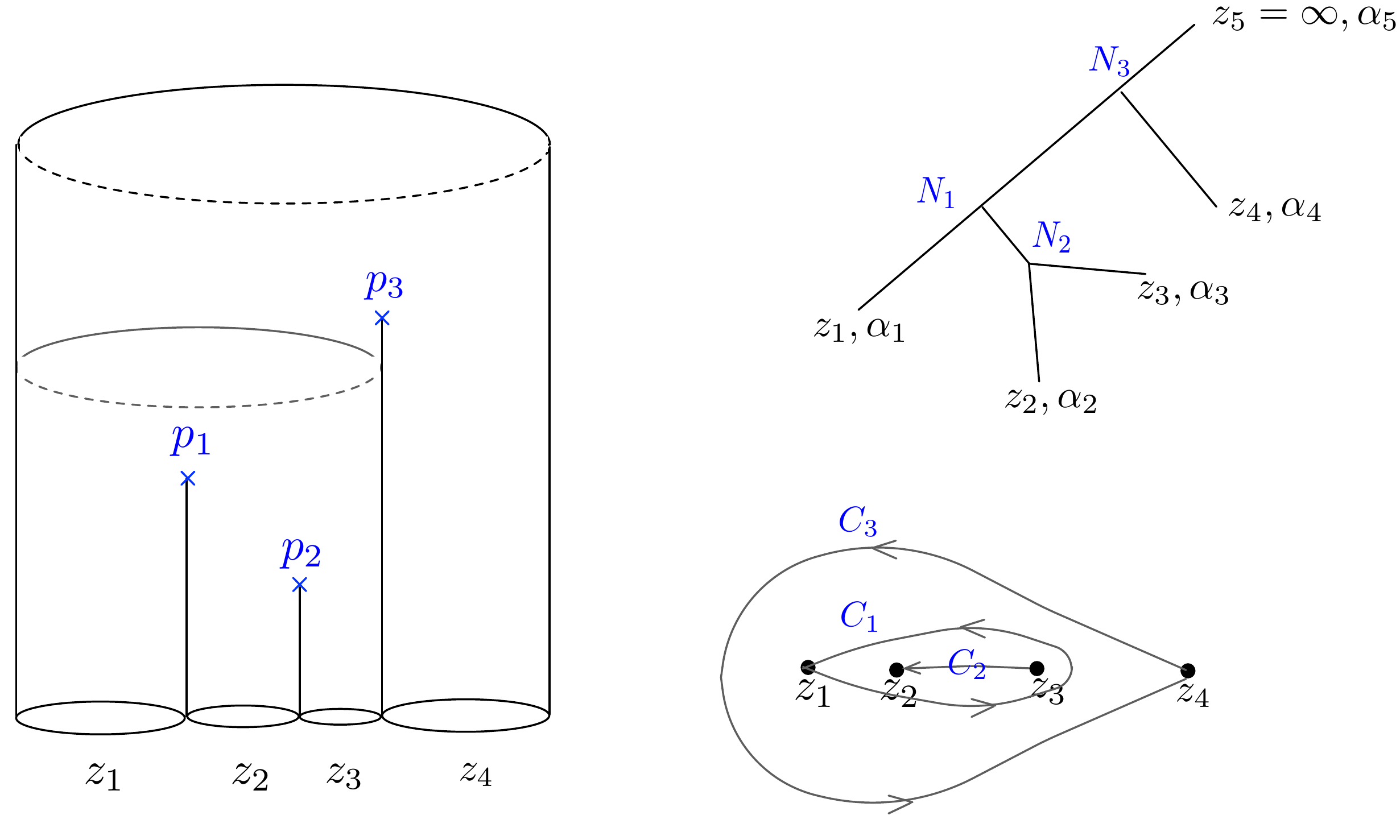} 
  \end{center}
    \caption{\it \footnotesize An example of the contours for computing a five-point conformal block. The ordering of the contour is given by the time-ordering of the interaction points on the light-cone diagram. Hence one should first perform the integration over the eigenvalue along the contours $C_2$, then $C_1$, and finally $C_3$. }
    \label{5-point_figure}
\end{figure}
The above conclusion can be extended to the cases when the straight line from a zero $p_\ell$ to a nearby pole $z_\ell$ cuts through a larger gluing curve connecting more than two tubes to the rest of the light-cone diagram. In this case the contour $C_\ell$ is a closed loop passing through $z_{\ell+1}$ and enclosing all the poles enclosed in the gluing curve. Finally, obviously the same result applies when we swap left and right and consider a situation where the segment from $p_\ell$ to a nearby pole $z_{\ell+1}$ cuts through a gluing curve. An example of such contours for computing the five-point conformal blocks is given in Figure \ref{5-point_figure}.

The final case to consider is when both of the line segments emanating from the zero $p_\ell$ to the pole $z_\ell$ and to $z_{\ell+1}$ cut through some gluing curves. Obviously, the corresponding Dehn twists on the left- and the right-hand side commute with each other and we can simply apply the above argument separately on them and combine the final results.

Moreover, we would like to bring the readers' attention to a further subtlety regarding the ordering of the contours. When considering the $\b$-ensemble deformation of the matrix model when $\b$ is not integral, an inspection of the integral formula (\ref{matrix_integral_beta}) shows that the phase $e^{\f_\ell}$ incurred by crossing the brach cut connecting the point $z_\ell$ to infinity might in fact depend on the pre-existing contours. Hence in this case the contours will no longer be simple tensor products of the same cycles but will have to be ordered instead. There is a completely natural way how this can be done. Namely, the time-ordering of the critical points on the light-cone diagram provides a natural ordering among distinct classes of contours corresponding to distinct critical points. We should hence perform the integration along the integration cycle $C_\ell$ corresponding to the earliest interaction point $p_\ell$, and move forward in time. See Figure \ref{5-point_figure} for an example. This ordering property of the contours should be regarded as a part of the definition for the non-perturbative blocks and will be illustrated in the examples discussed in section \ref{Examples}. 

Finally let us comment on the contours relevant for the comparison with the conformal blocks described in section \ref{The Liouville Conformal Blocks}. By definition, with fixed external momenta, conformal blocks in a given channel with different internal momenta furnish a diagonal basis for the monodromy transformation of the punctures on the Riemann surface that are equivalent to the Dehn twists in the corresponding light-cone diagram. Hence it is natural to conjecture that the matrix integral (\ref{matrix_integral_beta}) with the monodromy-diagonal contours described above, is {\it the same} as the conformal block described in section \ref{The Liouville Conformal Blocks} up to overall (leg) factors. In the next section we will show some examples of such equivalence.

\section{Examples}
\label{Examples}
\setcounter{equation}{0}
In the last section we have provided a prescription for the contours of the matrix integrals, and proposed its relation to the Liouville conformal blocks. 
In this section, we will illustrate and test our prescription by studying a few examples in detail. 

\subsection{Three-Point Functions}

\begin{figure}[h]
  \begin{center}
    \includegraphics[width=4in]{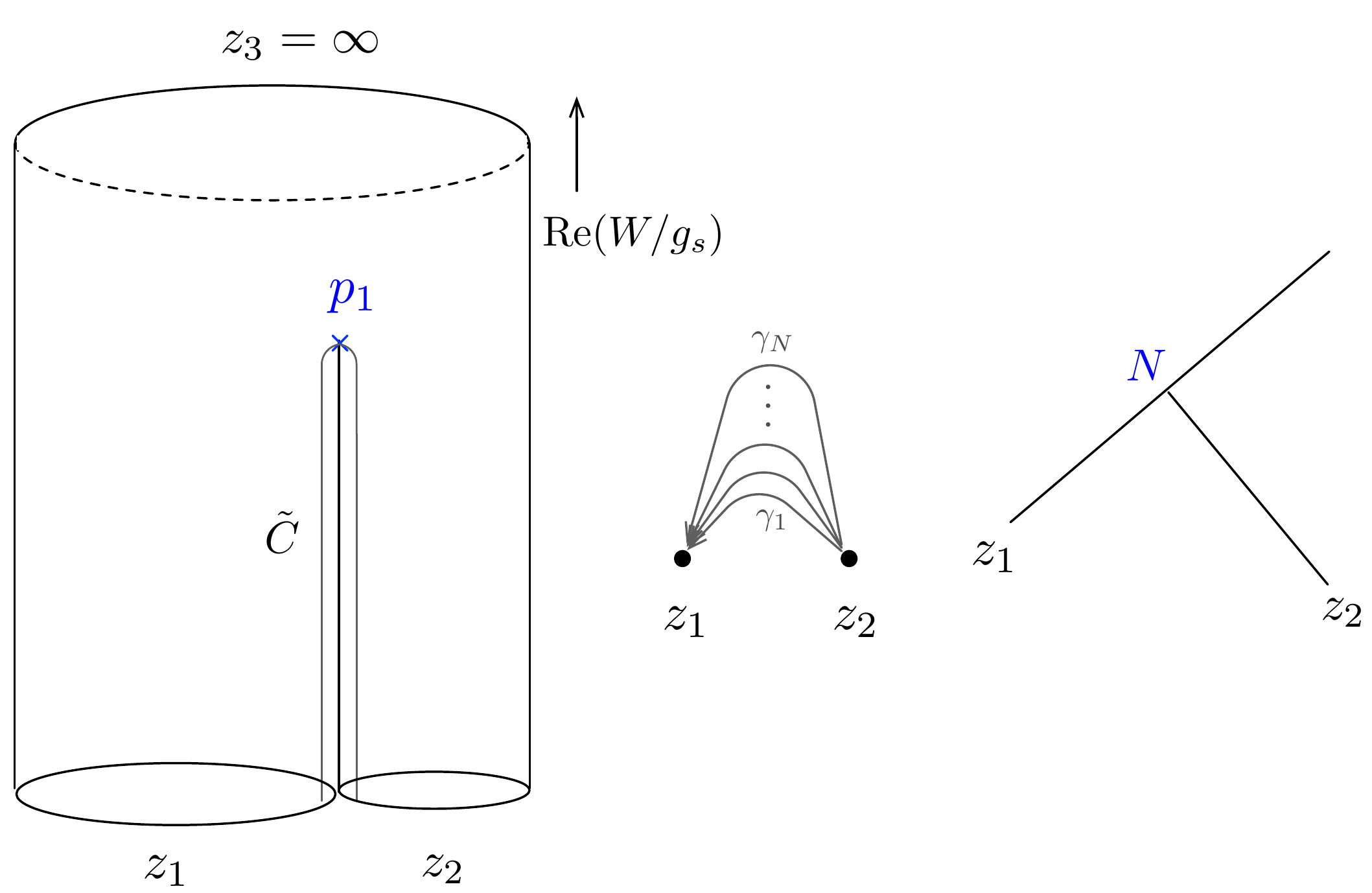} 
  \end{center}
    \caption{\it \footnotesize The simplest light-cone diagram is given by a single pair of pants. In this case the contour is simply $\til C$ given by the downward gradient flow of $\text{Re}(W/g_s)$ emanating from the unique critical point of the potential. The $N$ contour lines are ordered in such a way that they do not intersect with each other. The corresponding quantity on the 2d CFT side is the chiral half of  the three-point function, with the three momenta determined by the matrix potential $W$ and the rank of the matrix $N$.}
    \label{3point}
\end{figure}

 The simplest non-trivial example is given a pair of pants, which is the light-cone diagram corresponding to the potential 
 $$
{W(z)}/g_s ={m_1}\, \log(z-z_1)+{m_2}\, \log(z-z_2)\;.
 $$
 
According to the dictionary given in section \ref{The Liouville Conformal Blocks}, the $\b$-ensemble matrix model partition function of a rank $N$ matrix with the above potential corresponds to the chiral half of the three-point function, with the momenta of the inserted operators $V_{\a_1}(z_1)$, $V_{\a_2}(z_2)$ and $V_{\a_3}(z_3=\inf)$ given by
\be\label{3point_momenta}
\a_{1,2}= -\frac{1}{2b}m_{1,2}\;,\; \a_3 = Q- N b -\a_1-\a_2 \;.
\ee
To keep the notation uniform we also define $$ m_{3}=-{2b}\a_3 = -m_1-m_2-2(N-1)\b-2\;.$$

Now we shall apply the contour prescription described in previous sections to this simple case. As can be seen in Figure \ref{3point}, the contour given by the simple gradient line on the $W$-plane does not cut through any gluing curve in this case and is therefore free from any Dehn twist ambiguity. Therefore, we shall use the contour $\til C=[z_2,z_1]$ in our matrix integral. The discussion in section \ref{The Liouville Conformal Blocks} then suggests that the matrix integral
\bea\label{3point_integral} &&Z(z_1,z_2)=z_{21}^{m_1m_2/2\b}\times \\\notag
&& \int_{z_2}^{z_1}du_N\dotsi  \int_{z_2}^{z_1} du_1 \prod_{i=1}^N  (u_i-z_1)^{m_1}(u_i-z_2)^{m_2} \, \prod_{1\leq i<j\leq N} (u_i-u_j)^{2\b}
\eea
gives the chiral part of the three-point function up to normalization factors and the so-called leg-factors which only depend on individual momenta but not any combination of them. 


To see the validity of the above claim, first recall that the $SL(2,\C)$ invariance dictates the three-point functions to have the following form
$$
\langle V_{\a_3}(z_3)V_{\a_2}(z_2)V_{\a_1}(z_1)\rangle =C_{\a_1,\a_2,\a_3} \, |z_{21}|^{2\D_{21}} |z_{32}|^{2\D_{32}} |z_{13}|^{2\D_{13}}\;,
$$
where $\D_{ij} = \D_k -\D_i-\D_j $, $k\neq i, k \neq j$.
Hence all information of the three-point function is encoded in the object
$$
C_{\a_1,\a_2,\a_3}  = \lim_{z_3\to \inf} |z_3|^{-4\D_3} \langle V_{\a_3}(z_3)V_{\a_2}(0)V_{\a_1}(1)\rangle\;.
$$

On the other hand, a simple change of variables brings the matrix integral (\ref{3point_integral}) into the following form
\bea \notag Z(z_1,z_2) &=&z_{21}^{\D_{12}} \int_{0}^1 dt_N \dotsi \int_{0}^1dt_1  \prod_{i=1}^N t_i^{m_2}(t_i-1)^{m_1} \!\!\! \prod_{0\leq i<j\leq N} (t_i-t_j)^{2\b}\;.
\eea
Hence we see that the matrix integral $Z(z_1,z_2)$ indeed transforms in the correct way under $SL(2,\C)$ transformations that preserve $z_3 = \inf$. In particular, it has the correct transformation $Z(z_1,z_2) \to e^{2\p i \D_{21}} Z(z_1,z_2)$ under the monodromy transformation $z_{21}\to e^{2\p i }z_{21}$.
Furthermore, when fixing the location of the insertions using $SL(2,\C)$ transformation to be at $0,1,\inf$, we expect 
\bea \notag Z(1,0) &=& \int_{0}^1 dt_N \dotsi \int_{0}^1dt_1  \prod_{i=1}^N t_i^{m_2}(t_i-1)^{m_1} \!\!\! \prod_{0\leq i<j\leq N} (t_i-t_j)^{2\b}\;
\eea
to be a chiral half of the three-point constant $C_{\a_1,\a_2,\a_3}$, which is known as the DOZZ formula in the context of Liouville conformal theory \cite{Zamolodchikov1996,Dorn1994}.

The above integral is given by Selberg's formula, which reads
\bea\label{selberg}
Z(1,0)
 = (-1)^{m_1 N} N!\,\prod_{j=0}^{N-1} \frac{\G((j+1)\b)\G(1+m_1+j\b)\G(1+m_2+j\b)}{\G(\b)\G(-m_3-j\b)}.
\eea

To put Selberg's formula in a form closer to that of the the DOZZ formula, we use the following relation between the $\G$-function and the Double $\G$-function 
$$
\G(x) = \sqrt{2\p} \o_1^{1/2-x} \frac{\G_2(\o_1 x\lvert \o_1,\o_2)}{\G_2(\o_1 x+\o_2\lvert \o_1,\o_2)}\;.
$$
Moreover, eventually we would like to analytically continue the blocks obtained from matrix integrals to general values of momenta $\a_{1},\a_{2},\a_{3}$ not necessarily corresponding to integral units of the screening charges. To facilitate this, we would like to write the final answer in terms of the momenta $\a_{1,2,3}$ alone by eliminating the rank $N$ of the matrix from the final answer using the relation $N=(Q-\sum_{i=1}^3\a_i)/b$. By the same token we will now adopt a new notation $${\cal F}_{\a_1,\a_2,\a_3}=Z(1,0)\;.$$ Our proposed relation between the matrix integral and the conformal blocks then states that the DOZZ three-point function $C_{\a_1,\a_2,\a_3}$
is given by
$$
C_{\a_1,\a_2,\a_3}\sim {\cal F}_{\a_1,\a_2,\a_3}\, {\cal F}_{\a_1^\ast,\a_2^\ast,\a_3^\ast}\;,
$$
where ``$\sim$" signifies equality up to leg factors and overall normalization factors. 

Using the shorthand notation $\G_b(x)=\G_2(x\lvert b,b^{-1})$, we get
\bea\notag
{\cal F}_{\a_1,\a_2,\a_3} &\sim &\frac{\G_b(\a_1+\a_2+\a_3-Q)}{\G_b(0)}\frac{\G_b(\a_2+\a_3-\a_1)}{\G_b(Q-2\a_1)}\\ \notag
&& \qquad \times \ \  \frac{\G_b(\a_3+\a_1-\a_2)}{\G_b(Q-2\a_2)}  \frac{\G_b(Q+\a_3-\a_1-\a_2)}{\G_b(2\a_3)} \;,
\eea
which can be shown to satisfy the above relation. In fact, the leg factors can also be determined if we impose the following two simple conditions related to the two-point function: 
$$
\lim_{\e\to 0} C_{\a,\e,Q-\a}  =\lim_{\e\to 0} C_{\e,\a,Q-\a}\sim \frac{1}{\e} \quad,\quad \lim_{\e\to 0} C_{\a,\e,\a}  =\lim_{\e\to 0} C_{\e,\a,\a}\sim \frac{1}{\e}S(\a)\;,
$$
where $S(\a)$ is the so-called reflection amplitude given by
$$
S(\a) \sim \frac{\G_b(2Q-2\a)\G_b(2\a-Q)}{\G_b(2\a)\G_b(Q-2\a)}\;.
$$

From the above requirement, we conclude that the leg factors are given by
$$
f(\a) = \frac{\G_b(2\a-Q)}{\G_b(2\a)} 
$$
for the incoming strings in our light-cone diagram and $g(\a)=\frac{1}{f(\a^\ast)}$ for the outgoing ones. 
Incorporating these leg factors, we finally get 
\bea\notag
&&\frac{f(\a_1)f(\a_2)}{f(\a^\ast_3)}  \,{\cal F}_{\a_1,\a_2,\a_3}\, {\cal F}_{\a_1^\ast,\a_2^\ast,\a_3^\ast} \\\notag
&&\sim \frac{\U'_0 \U(2\a_1)\U(2\a_2)\U(2\a_3)}{\U( \a_1+\a_2+\a_3 -Q)\U(\a_2+\a_3-\a_1)\U(\a_3+\a_1-\a_2)\U(\a_1+\a_2-\a_3)}\;,
\eea
where $$\U(x)= \frac{1}{\G_b(x)\G_b(Q-x)}\;.$$ 
This is precisely the DOZZ formula \cite{Zamolodchikov1996,Dorn1994} apart from the extra normalization factor 
$$
\left(\p \m \frac{\G(b^2)}{\G(1-b^2)}b^{2-2b^2}\right)^{(Q-\sum_{\ell=1}^3\a_\ell)/b}\;.
$$
Notice that this normalization factor depends on the coefficient $\m$ of the exponential term $\int d^2z\, \m \,e^{2b\varphi}$ in the Liouville action which can be shifted away by shifting $\f\to \f+$ const, and is therefore clearly scheme-depedent. Hence, we see that in the case of the three-point function, the matrix model indeed captures the chiral information. The relation between the matrix integral and the DOZZ formula was also discussed in \cite{Schiappa2009}. 

Finally, note that our matrix integral 
$${\cal F}_{\a_1,\a_2,\a_3} \sim\,\prod_{j=0}^{N-1} \frac{\G((j+1)\b)\G(1+m_1+j\b)\G(1+m_2+j\b)}{\G(\b)\G(-m_3-j\b)}
$$ 
is not the only possible chiral half of the DOZZ formula $C_{\a_1,\a_2,\a_3}$. There are a few discrete choices that can be made here. Obviously, some other possibilities are given by permuting $\a_1,\a_2,\a_3$ in the above formula. More generally, it is not hard to see that by replacing the factor $\G(1+m_i+j\b)$ with $1/\G(-m_i-j\b)$ in the product or another way around, we obtain another object which squares to $C_{\a_1,\a_2,\a_3}$. In particular, another possibility is given by 
$${\cal \hat F}_{\a_1,\a_2,\a_3} \sim\,\prod_{j=0}^{N-1} \frac{\G((j+1)\b)\G(1+m_1+j\b)}{\G(\b)\G(-m_2-j\b)\G(-m_3-j\b)}\;,\; $$
satisfying
\be \label{3point_2}
{\cal \hat F}_{\a_1,\a_2,\a_3}\,{\cal \hat F}_{\a_1^\ast,\a_2^\ast,\a_3^\ast} \sim C_{\a_1,\a_2,\a_3}\;.
\ee
This property of the three-point function will be useful later in our discussion of the four-point conformal blocks. 

\begin{figure}
  \begin{center}
    \includegraphics[width=4.5in]{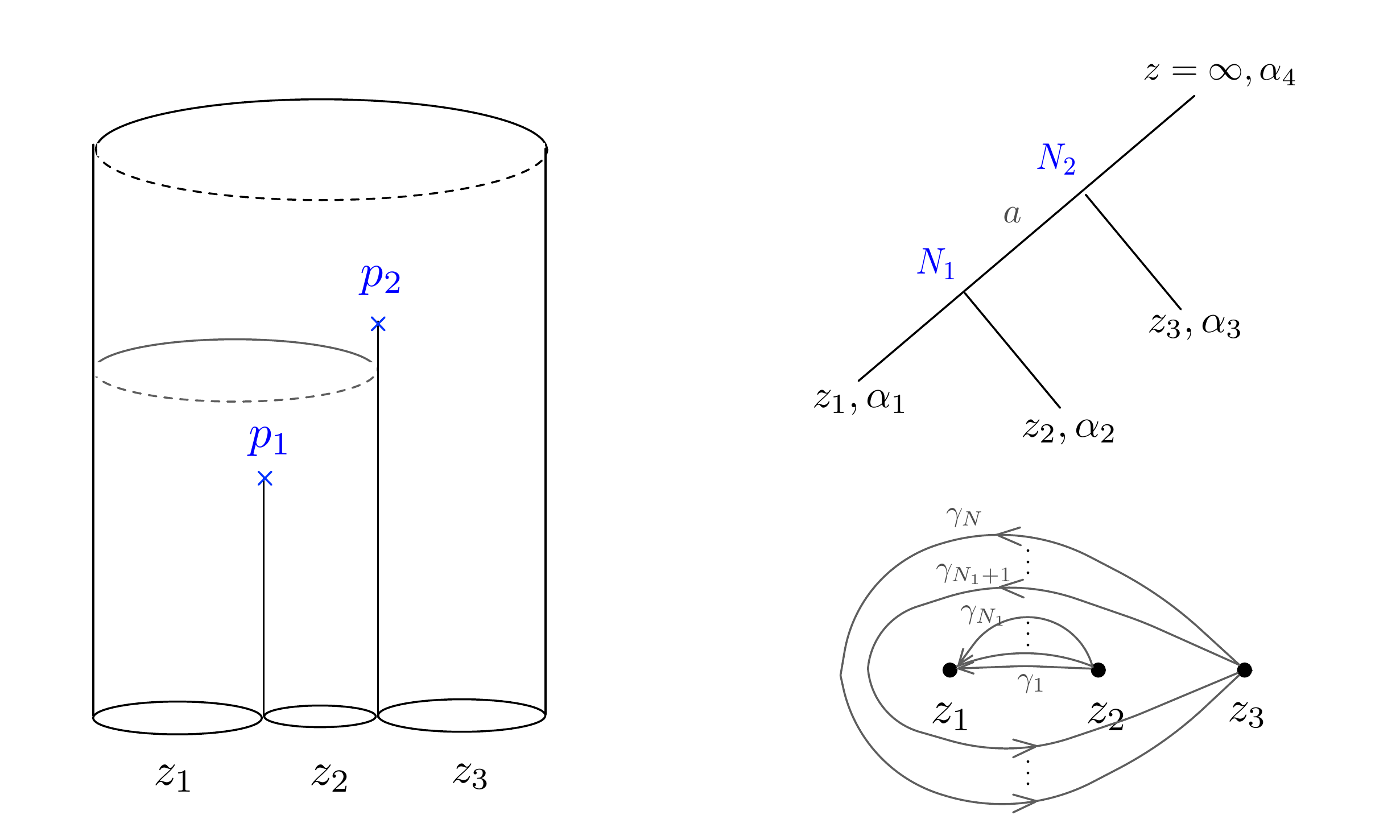} 
  \end{center}
    \caption{\it \footnotesize On the left is a light-cone diagram corresponding to the potential (\ref{4point_potential}). Together with the eigenvalue distribution $N_1$, $N_2\!=\!N-N_1$, it maps to the $s$-channel conformal block shown on the right, with the internal momenta given by $a=\a_1+\a_2+N_1 b$. }
    \label{4point_2_FIG}
\end{figure}

\subsection{Four-Point Functions}

Now we consider the four-point function corresponding to the following matrix potential 
 \be\label{4point_potential}
{W(z)}/g_s ={m_1}\, \log(z-z_1)+{m_2}\, \log(z-z_2)+{m_3}\, \log(z-z_3)\;.
 \ee
Depending on the parameters $m_{1,2,3}$ and $z_{1,2,3}$, the corresponding light-cone diagram might take a different shape corresponding to a different pair of pants decomposition. In Figure \ref{4point_2_FIG} we show one of the possibilities which corresponds to the $s$-channel conformal blocks. According to the prescription in section \ref{The Conformal Blocks versus the Topological String Blocks}, the corresponding contour of integration is the one shown in  Figure \ref{4point_2_FIG}.

The $SL(2,\C)$ invariance dictates the following form for the four-point function
\bea\notag
&&\langle V_{\a_4}(z_4)V_{\a_3}(z_3)V_{\a_2}(z_2)V_{\a_1}(z_1)\rangle \\\notag&& \ \ \ =  |z_{43}|^{2(\D_{1}+\D_2-\D_3-\D_4)} |z_{42}|^{-4\D_{2}} |z_{41}|^{2(\D_{3}+\D_2-\D_1-\D_4)}|z_{31}|^{2(\D_{4}-\D_1-\D_2-\D_3)}\\ \notag
& & \qquad \qquad \times \ \ G_{\a_1,\a_2,\a_3,\a_4}(z,\bar z)
\eea
where $z_{ij}=z_j -z_i$ and $z$ is the cross ratio of the four points
$$z=\frac{z_{43}z_{21}}{z_{42}z_{31}} \;.$$ 

In particular, all the information of interest is contained in the following object
$$
G_{\a_1,\a_2,\a_3,\a_4}(z,\bar z)  = \lim_{z_4\to \inf} |z_4|^{4\D_4} \langle V_{\a_4}(z_4)V_{\a_3}(1)V_{\a_2}(z)V_{\a_1}(0)\rangle\;.
$$

Moreover, from the $s$-channel sewing procedure, using the operator algebra one can see that it has the following decomposition into conformal blocks with different internal momenta $a$
$$
G_{\a_1,\a_2,\a_3,\a_4}(z,\bar z)  = \sum_\a C_{\a_1,\a_2,a} C_{a^\ast,\a_3,\a_4} \, \left\vert{\cal F}^s_{a}\!\left[\!\!\begin{array}{cc} \a_1&\a_2\\ \a_3&\a_4\end{array};z\right] \right\vert^2 \;,
$$
where the prefactors, given in terms of the DOZZ three-point functions, are chosen such that the conformal blocks have the following normalization
$${\cal F}^s_{a}\!\left[\!\!\begin{array}{cc} \a_1&\a_2\\ \a_3&\a_4\end{array};z=0\right] =1\;.
$$

In this subsection we would like to verify the relation between the matrix integrals and the above four-point conformal blocks. First we would like to see that the matrix integrals indeed take the above form. From the discussion in section \ref{The Liouville Conformal Blocks}, we expect the four-point conformal block with operator insertions at locations $z_1,z_2,z_3,z_4=\inf$ to be given by the matrix integral
\bea\notag
&&\!\!\!\!\!\!\prod_{1\leq k < \ell \leq 3} (z_{\ell}-z_k)^{m_\ell m_k/2\b} \idotsint \prod_{i=1}^N du_i   \prod_{1\leq i < j \leq N} (u_j-u_i)^{2\b}\,\prod_{i=1}^N\prod_{\ell=1}^{3} (u_i-z_\ell)^{m_\ell}\\\notag
&& \ \ = z_{13}^{\D_4-\D_1-\D_2-\D_3} z^{m_1m_2/2\b}(1-z)^{m_2m_3/2\b} \\\notag && \qquad \times\idotsint \prod_{i=1}^N du_i \,u_i^{m_1} (u_i-z)^{m_2} (u_i-1)^{m_3}  \!\!\! \prod_{1\leq i < j \leq N} (u_j-u_i)^{2\b} \;,
\eea
 where we have used the identification $\sum_{\ell=1}^4\a_\ell = Q-Nb$ and $z=z_{21}/z_{31}$. 

Hence we see that the matrix integrals transform correctly under $SL(2,\C)$ transformations which preserve $z_4=\inf$. From the discussions in section \ref{The Liouville Conformal Blocks} and section \ref{The Conformal Blocks versus the Topological String Blocks}, we then expect the matrix model integral with the contour $\hat C_a$ corresponding to the given internal momenta $a$ to be related to the conformal block as
\bea\notag
Z_a(z)&=&z^{m_1m_2/2\b}(1-z)^{m_2m_3/2\b} \\ \notag
& & \qquad \times \ \int_{\hat C_a} \prod_{i=1}^N du_i \,u_i^{m_1} (u_i-z)^{m_2} (u_i-1)^{m_3}  \prod_{1\leq i < j \leq N} (u_j-u_i)^{2\b} \\ \label{4point_1} &=&  f_{\a_1,\a_2,a}\, f_{a^\ast,\a_3,\a_4} \, {\cal F}^s_{a}\!\left[\!\!\begin{array}{cc} \a_1&\a_2\\ \a_3&\a_4\end{array};z\right] \;,
\eea
where $f_{\a_1,\a_2,a}$ is a chiral half of the DOZZ three-point function, namely $$f_{\a_1,\a_2,a} f_{\a_1^\ast,\a_2^\ast,a^\ast} \sim C_{\a_1,\a_2,a}$$ up to leg factors and overall normalization factors. In appendix \ref{Degenerate Four-Point Conformal Blocks} we will show the validity of this relation explicitly for the special case when one of the insertions carry a degenerate momentum.

\section{Discussions}
\label{discussions}

In this paper we defined a non-perturbative completion of the topological string theories with a dual open string description. This completion involved discrete choices corresponding to supersymmetric boundary conditions in the physical theory. We then applied this definition to the topological string theories capturing the Nekrasov partition functions of $SU(2)$ ${\cal N}=2, d=4$ superconformal gauge theories that are related to the 2d conformal blocks of the Liouville CFT through the so-called AGT correspondence. An interesting feature of the map between topological string blocks and conformal blocks is that to relate to the latter we are forced to consider blocks that span the eigenbasis of various monodromy operations. Also from the topological string point of it seems natural to consider such a basis. It would hence be interesting to investigate the physical meaning of such eigenbasis for various monodromy operations of the parameters in the dual matrix models for more general topological string theories. 

The results in the present paper admit several generalizations. First, similar treatments should apply to the more general cases of 2d Toda CFT's, corresponding to quiver matrix models on one side and $4d$ gauge theories with gauge groups different from $SU(2)$ on the other side. Also the higher genus conformal blocks should have an integral representation with an analogous interpretation in topological string theory. Finally, it would be desirable to make more precise the analytic continuation of the eigenvalue multiplicities $N_\ell$ and the mass parameters $m_\ell$ that is implicit in our paper. We expect the former to be very similar to the analytic continuation done in \cite{Teschner2001}.

Finally, the prominent role played by the closed string diagram in the light-cone gauge in our analysis is rather intriguing. In particular, in the case of the $SU(2)$ theory on $S^4$, when the two $\O$-background parameters $\e_1$ and $\e_2$ are equal and inversely proportional to the radius of the sphere \cite{Pestun2007}, the Liouville theory has $b^2=1$ and hence central charge $c=25$. Considering the $U(2)$ theory and incorporating the $U(1)$ factor we then get the total central charge $c=26$, corresponding to $d=1$ critical bosonic string on the nose!
Even though the light-cone diagrams and the fields living on them correspond exactly to those of critical bosonic string,
there is one important difference:  In the gauge theory context we are not instructed to integrate over the lengths and twists
of the internal tubes of the light-cone diagram.  These would have corresponded  in the gauge theory setup to integrating over the
gauge coupling constants. It is therefore natural to expect that the string theory quantities correspond to making various parameters of the four-dimensional gauge theories dynamical. It will be interesting to develop this connection further. 

\section*{Acknowledgments}
We would like to thank J. Teschner for valuable discussions. M.C. would like to thank the University of Amsterdam, LPTHE Jussieu and LPTENS in Paris, Max Planck Institut f\"ur Physik in Munich, and KITP Santa Barbara, for kind hospitality during the process of this project. R.D. and C.V. also thank the Simons Workshop in Mathematics and Physics 2010 for providing a stimulating research environment as well as for its warm hospitality. The research of M.C. is supported in part by DOE grant DE-FG02-91ER40654. The research of R.D. was supported by a NWO Spinoza grant and the FOM program {\it String Theory and Quantum Gravity}. The research of C.V. was supported in part by NSF grant PHY-0244821. 
\appendix

\section{Degenerate Four-Point Conformal Blocks}
\label{Degenerate Four-Point Conformal Blocks}
\setcounter{equation}{0}
In this appendix we explicitly show that the expected relation between the four-point conformal blocks and the matrix model blocks (\ref{4point_1}) indeed holds for the special case with degenerate insertions. When one of the insertions has degenerate momenta $\a= -nb/2 -m/2b$ for non-negative integers $n$ and $m$, the corresponding Virasoro representations are reducible and the four-point functions satisfy linear differential equations \cite{Belavin1984}. We will now focus on the simplest case with $\a_2 = -b/2$. In this case, the differential equation dictates the conformal blocks to be given in terms of the hypergeometric functions $_2F_1\big[\!\!\begin{array}{c}A,\,B\\C\end{array};z \big]$ as \cite{Belavin1984}
\bea\notag
{\cal F}^s_{a=\a_1-b/2}\!\!\left[\!\!\begin{array}{cc} \a_1&-b/2\\ \a_3&\a_4\end{array};z\right] &\!\!= \!\!&  z^{\D_a-\D_1-\D_2} (1-z)^{\D_{\a_2+\a_3}-\D_3-\D_2} \\ \notag&&\times \  _2F_1
\!\Big[\!\!\begin{array}{c} N\beta,1+m_4+(N-1)\beta\\ -m_1+\beta\end{array};z\Big] 
\\ \label{known_degenerate_4point}
{\cal F}^s_{a=\a_1+b/2}\!\!\left[\!\!\begin{array}{cc} \a_1&-b/2\\ \a_3&\a_4\end{array};z\right] &\!\!=\!\!&z^{\D_a-\D_1-\D_2} (1-z)^{\D_{\a_2+\a_3}-\D_3-\D_2}\\ \notag&&\times \ _2F_1\! \Big[\!\!\begin{array}{c} 1+m_1+(N-1)\beta, -m_3-(N-1)\b\\ 2+m_1-\beta\end{array};z \Big]
\eea
and all other conformal blocks with internal momentum $a \neq \a_1 \pm b/2$ vanish identically. The hypergeometric function $_2F_1(A,B;C;z)$ is analytic at the origin $z=0$. More precisely, it is normalized such that $$_2F_1\! \Big[\!\!\begin{array}{c}A,\,B\\C\end{array};z=0 \Big]=1\;.$$ 
Together with $$z^{1-C} \;_2F_1\!\Big[\!\!\begin{array}{c}A+1-C,B+1-C\\2-C\end{array};z \Big]$$ it spans the space of solutions to the following ODE
\be\label{hyper_eqn}
z(1-z) f'' + [C- (1+A+B)z] f' - A B f=0\;.
\ee

Now we would like to compare the matrix integral with the above known result. In the present case, we have $m_2=-\b$ and the relevant matrix model quantity (\ref{4point_1}) is 
$$
z^{-m_1/2}(1-z)^{-m_3/2} \int_{\hat C_a} \prod_{i=1}^N du_i \,u_i^{m_1} (u_i-z)^{-\b} (u_i-1)^{m_3}  \prod_{1\leq i < j \leq N} (u_j-u_i)^{2\b} \;,
$$
where the contour $\hat C_a$ is shown in Figure \ref{4point_2_FIG}, with now $z_1=0,z_2=0,z_3=1$ and $a = \a_1+(N_1-1)\b$.

The above integral is exactly the kind which has been discussed in \cite{kaneko}, where it was shown that the above integral, no matter what the contour is, satisfies the hypergeometric equation (\ref{hyper_eqn}) with 
$$ A=N\b, \;B= m_4+1+(N-1)\b, \;C= -m_1+\b\;.
$$
This fact together with the boundary condition near $z=0$, which depends on the contour, is then sufficient to determine the integral. 

To see this, first consider the contour $\hat C_{\a=\a_1-b/2}$ depicted in Figure \ref{4point_2_FIG} with $N_1\!=\!0, N_2\!=\!N$. Since there are no other punctures on the sphere, the contour that passes through $z_3\!=\!1$ and circles $z_1\!=\!0$ and $z_2\!=\!z$ can be deformed to a contour passing through $z_3$ and circling $z_4=\inf$. Hence we obtain 
\bea\notag
&&\int_{\hat C_{a=\a_1-b/2}} \prod_{i=1}^N du_i \,u_i^{m_1} (u_i-z)^{-\b} (u_i-1)^{m_3}  \prod_{1\leq i < j \leq N} (u_j-u_i)^{2\b} \\ \notag
&\sim&  \prod_{j=0}^{N-1} \left(1-e^{2\p i (m_4+j\b)}\right)  \\ \notag&&\times  \int_1^\inf\dotsi\int_1^\inf
\prod_{i=1}^N du_i \,u_i^{m_1} (u_i-z)^{-\b} (u_i-1)^{m_3}  
\prod_{1\leq i < j \leq N} (u_j-u_i)^{2\b} \\ \notag
&\sim& \prod_{j=0}^{N-1}  \frac{1}{\G(1+m_4+j\b)\G(-m_4-j\b)}  \\ \notag&&\times \int_0^1 \dotsi\int_0^1 \prod_{i=1}^N d\l_i \,\l_i^{m_4} (1-z\l_i)^{-\b} (\l_i-1)^{m_3}  
\prod_{1\leq i < j \leq N} (\l_j-\l_i)^{2\b}
\eea
where we have used 
$$
1-e^{2\p i x} = \frac{ -2 i \p e^{\p i x}}{\G(x)\G(1-x)}
$$
in the last equation. Notice that the integral in the last line is analytic when $z\to 0$ and in particular is given by 
$$
{\cal F}_{\a_3,\a_4,\a_1-b/2} = \int_0^1 \dotsi\int_0^1 \prod_{i=1}^N d\l_i \,\l_i^{m_4} (\l_i-1)^{m_3}  
\prod_{1\leq i < j \leq N} (\l_j-\l_i)^{2\b}\;,
$$ 
we then get 
\bea\notag
Z_{a=\a_1-b/2}(z)\!\! &=&\!\!z^{-m_1/2}(1-z)^{-m_3/2} \\ \notag&&\times \int_{\hat C_{\a=\a_1-b/2}} \prod_{i=1}^N du_i \,u_i^{m_1} (u_i-z)^{-\b} (u_i-1)^{m_3}  \prod_{1\leq i < j \leq N} (u_j-u_i)^{2\b}\\ \notag
&=& \!\!z^{-m_1/2}(1-z)^{-m_3/2} \prod_{j=0}^{N-1}  \frac{1}{\G(1+m_4+j\b)\G(-m_4-j\b)}  \; \\\notag &&\times\;\;{\cal F}_{\a_3,\a_4,\a_1-b/2} \cdot\ _2F_1
\!\Big[\!\!\begin{array}{c}N\beta,1+m_4+(N-1)\beta\\-m_1+\beta\end{array};z\Big]\\ \notag
&=&\!\! z^{-m_1/2}(1-z)^{-m_3/2} {\cal \hat F}_{\a_3,\a_4,\a_1-b/2}\;\cdot\;_2F_1\!\Big[\!\!\begin{array}{c}N\beta,1+m_4+(N-1)\beta\\-m_1+\beta\end{array};z\Big]\;.
\eea

Let us now consider a second contour  $\hat C_{a=\a_1+b/2}$ depicted in Figure \ref{4point_2_FIG} with $N_1\!=\!1, N_2\!=\!N-1$. A similar calculation shows
\bea\notag
\!\!&&\int_{\hat C_{a=\a_1+b/2}} \prod_{i=1}^N du_i \,u_i^{m_1} (u_i-z)^{-\b} (u_i-1)^{m_3}  \prod_{1\leq i < j \leq N} (u_j-u_i)^{2\b}\\\notag
\!\!&\sim&   \prod_{j=0}^{N-2} \left(1-e^{2\p i (m_4+j\b)}\right)\int_1^\inf\dotsi\int_1^\inf
\prod_{i=1}^{N-1} du_i \int_0^z du \Big\{\,u^{m_1} (u-z)^{-\b} (u-1)^{m_3}   \\\notag && \times \prod_{i=1}^{N-1} u_i^{m_1} (u_i-z)^{-\b} (u_i-1)^{m_3} (u-u_i)^{2\b} \!\!
\prod_{1\leq i < j \leq N-1} \!\!\!(u_j-u_i)^{2\b}\Big\} \\ \notag
\!\!& \sim& z^{1+m_1-\b}\prod_{j=0}^{N-2}  \frac{1}{\G(1+m_4+j\b)\G(-m_4-j\b)}  \int_0^1\dotsi\int_0^1 \prod_{i=1}^{N-1} d\l_i \,\int_0^1 du\\ \notag&&\Big\{\,t^{m_1} (t-1)^{-\b} (zt-1)^{m_3}  \prod_{i=1}^{N-1}   \l_i^{m_4} (1-z\l_i)^{-\b} (1-\l_i)^{m_3}  (z u\l_i-1)^{2\b}\!\!\!\!
\prod_{1\leq i < j \leq N-1}\!\!\! (\l_i-\l_j)^{2\b}\Big\}\;.
\eea 
Notice again that the final integral is analytic at the origin $z=0$. Comparing with the basis of the solution to the hypergeometric equation we hence conclude
\bea\notag
Z_{a=\a_1+b/2}(z)&\!\!=\!\!&z^{-m_1/2}(1-z)^{-m_3/2}\\\notag &&\times\int_{\hat C_{\a=\a_1+b/2}} \prod_{i=1}^N du_i \,u_i^{m_1} (u_i-z)^{-\b} (u_i-1)^{m_3}  \prod_{1\leq i < j \leq N} (u_j-u_i)^{2\b}\\ \notag
&\!\!=& z^{1+m_1/2-\b}(1-z)^{-m_3/2} {\cal F}_{\a_1,-b/2,Q-\a_1-b/2} \,{\cal \hat F}_{\a_3,\a_4,\a_1+b/2} \\
&&\times \;_2F_1\Big[\!\!\begin{array}{c} 1+m_1+(N-1)\b,-m_3-(N-1)\b\\ 2+m_1-\b \end{array};z \Big]\;.
\eea

Similarly, by comparing the analyticity property of the integral near $z=0$ with the basis of solutions to the hypergeometric equation that the integral has to satisfy, we conclude that all other $Z_{a}=0$. Therefore, comparing the above results with (\ref{3point_2}) and (\ref{known_degenerate_4point}), we see that the proposed relation between the matrix model blocks and the conformal blocks (\ref{4point_1}) indeed holds for the case when one of the insertions has the degenerate momentum $\a=-b/2$.

\bibliography{ref_final.bib}{}

\providecommand{\href}[2]{#2}\begingroup\raggedright\begin{thebibliography}{10}

\bibitem{Gopakumar:1998vy}
R.~Gopakumar and C.~Vafa, ``{Topological Gravity as Large N Topological Gauge
  Theory},'' {\em Adv.Theor.Math.Phys.} {\bf 2,1998}
  (Adv.Theor.Math.Phys.2:413-442,1998)  413--442,
  \href{http://arxiv.org/abs/hep-th/9802016}{{\tt hep-th/9802016}}.

\bibitem{Gopakumar1999}
R.~Gopakumar and C.~Vafa, ``{On the Gauge Theory/Geometry Correspondence},''
  {\em Adv.Theor.Math.Phys.} {\bf 3} (1999)  1415--1443,
  \href{http://arxiv.org/abs/hep-th/9811131}{{\tt hep-th/9811131}}.

\bibitem{Dijkgraaf2002}
R.~Dijkgraaf and C.~Vafa, ``{Matrix Models, Topological Strings, and
  Supersymmetric Gauge Theories},'' {\em Nucl.Phys. B} {\bf 644} (2002)  3--20,
  \href{http://arxiv.org/abs/hep-th/0206255}{{\tt hep-th/0206255}}.

\bibitem{David1991}
F.~David, ``{Phases of the large N matrix model and nonperturbative effects in
  2-d gravity},''
\href{http://dx.doi.org/10.1016/0550-3213(91)90202-9}{{\em Nucl. Phys.} {\bf
  B348} (1991)  507--524}.

\bibitem{David1993}
F.~David, ``{Non-Perturbative Effects in Matrix Models and Vacua of Two
  Dimensional Gravity},'' {\em Phys.Lett. B} {\bf 302} (1993)  403--410,
  \href{http://arxiv.org/abs/hep-th/9212106}{{\tt hep-th/9212106}}.

\bibitem{Dimofte2009}
T.~Dimofte, S.~Gukov, J.~Lenells, and D.~Zagier, ``Exact results for
  perturbative chern-simons theory with complex gauge group,'' {\em
  Commun.Num.Theor.Phys.} {\bf 3:363-443,2009} (2009)  .

\bibitem{Witten2010}
E.~Witten, ``{Analytic Continuation Of Chern-Simons Theory},''
  \href{http://arxiv.org/abs/1001.2933}{{\tt 1001.2933}}.

\bibitem{Witten2010a}
E.~Witten, ``{A New Look At The Path Integral Of Quantum Mechanics},''
  \href{http://arxiv.org/abs/1009.6032}{{\tt 1009.6032}}.

\bibitem{Alday2009}
L.~F. Alday, D.~Gaiotto, and Y.~Tachikawa, ``{Liouville Correlation Functions
  from Four-dimensional Gauge Theories},''
  \href{http://arxiv.org/abs/0906.3219}{{\tt 0906.3219}}.

\bibitem{Wyllard2009}
N.~Wyllard, ``{$A_{N-1}$ conformal Toda field theory correlation functions from
  conformal N=2 SU(N) quiver gauge theories},''{\em JHEP} {\bf 0911:002,2009}
  (July, 2009)  , \href{http://arxiv.org/abs/0907.2189}{{\tt 0907.2189}}.

\bibitem{Dijkgraaf2009}
R.~Dijkgraaf and C.~Vafa, ``{Toda Theories, Matrix Models, Topological Strings,
  and N=2 Gauge Systems},'' \href{http://arxiv.org/abs/0909.2453}{{\tt
  0909.2453}}.

\bibitem{Schiappa2009}
R.~Schiappa and N.~Wyllard, ``{An $A_r$ threesome: Matrix models, 2d CFTs and
  4d N=2 gauge theories},'' \href{http://arxiv.org/abs/0911.5337}{{\tt
  0911.5337}}.

\bibitem{Mironov2010}
A.~Mironov, A.~Morozov, and S.~Shakirov, ``{Conformal blocks as Dotsenko-Fateev
  Integral Discriminants},'' \href{http://arxiv.org/abs/1001.0563}{{\tt
  1001.0563}}.

\bibitem{Mironov2010a}
A.~Mironov, A.~Morozov, and A.~Morozov, ``{Matrix model version of AGT
  conjecture and generalized Selberg integrals},''
  \href{http://arxiv.org/abs/1003.5752}{{\tt 1003.5752}}.

\bibitem{Morozov2010}
A.~Morozov and S.~Shakirov, ``{The matrix model version of AGT conjecture and
  CIV-DV prepotential},'' \href{http://arxiv.org/abs/1004.2917}{{\tt
  1004.2917}}.

\bibitem{Maruyoshi:2010pw}
K.~Maruyoshi and F.~Yagi, ``{Seiberg-Witten curve via generalized matrix
  model},'' \href{http://arxiv.org/abs/1009.5553}{{\tt 1009.5553}}.

\bibitem{Mironov:2010su}
A.~Mironov, A.~Morozov, and S.~Shakirov, ``{On Dotsenko-Fateev representation
  of the toric conformal blocks},'' \href{http://arxiv.org/abs/1010.1734}{{\tt
  1010.1734}}.

\bibitem{Eguchi2009}
T.~Eguchi and K.~Maruyoshi, ``{Penner Type Matrix Model and Seiberg-Witten
  Theory},'' \href{http://arxiv.org/abs/0911.4797}{{\tt 0911.4797}}.

\bibitem{Eguchi2010}
T.~Eguchi and K.~Maruyoshi, ``{Seiberg-Witten theory, matrix model and AGT
  relation},'' \href{http://arxiv.org/abs/1006.0828}{{\tt 1006.0828}}.

\bibitem{Dotsenko1985}
V.~S. Dotsenko and V.~A. Fateev, ``{Four Point Correlation Functions and the
  Operator Algebra in the Two-Dimensional Conformal Invariant Theories with the
  Central Charge c $<$ 1},''
\href{http://dx.doi.org/10.1016/S0550-3213(85)80004-3}{{\em Nucl. Phys.} {\bf
  B251} (1985)  691}.

\bibitem{Felder1989}
G.~Felder, ``{BRST Approach to Minimal Methods},''
\href{http://dx.doi.org/10.1016/0550-3213(89)90568-3}{{\em Nucl. Phys.} {\bf
  B317} (1989)  215}.

\bibitem{Dijkgraaf2007a}
R.~Dijkgraaf, C.~Vafa, and E.~Verlinde, ``{M-theory and a Topological String
  Duality},'' \href{http://arxiv.org/abs/hep-th/0602087}{{\tt hep-th/0602087}}.

\bibitem{DSV-unp}
R.~Dijkgraaf, P.~Sulkowski, and C.~Vafa, ``unpublished,''.

\bibitem{Cecotti2010}
S.~Cecotti, A.~Neitzke, and C.~Vafa, ``{R-Twisting and 4d/2d
  Correspondences},'' \href{http://arxiv.org/abs/1006.3435}{{\tt 1006.3435}}.

\bibitem{Aganagic2009}
M.~Aganagic and M.~Yamazaki, ``{Open BPS Wall Crossing and M-theory},''{\em
  Nucl.Phys.B} {\bf 834:258-272,2010} (Nov., 2009)  ,
  \href{http://arxiv.org/abs/0911.5342}{{\tt 0911.5342}}.

\bibitem{WarnerNucl.Phys.B450:663-6941995}
N.~P. Warner, ``{Supersymmetry in Boundary Integrable Models},'' {\em
  Nucl.Phys.B} {\bf 450:663-694,1995} (Nucl.Phys.B450:663-694,1995)  ,
  \href{http://arxiv.org/abs/hep-th/9506064}{{\tt hep-th/9506064}}.

\bibitem{Hori2007}
K.~Hori, A.~Iqbal, and C.~Vafa, ``{D-Branes And Mirror Symmetry},''
  \href{http://arxiv.org/abs/hep-th/0005247}{{\tt hep-th/0005247}}.

\bibitem{Marino2007a}
M.~Marino, R.~Schiappa, and M.~Weiss, ``{Nonperturbative Effects and the
  Large-Order Behavior of Matrix Models and Topological Strings},''
  \href{http://arxiv.org/abs/0711.1954}{{\tt 0711.1954}}.

\bibitem{Marino2008}
M.~Marino, ``{Nonperturbative effects and nonperturbative definitions in matrix
  models and topological strings},''{\em JHEP} {\bf 0812:114,2008} (May, 2008)
  , \href{http://arxiv.org/abs/0805.3033}{{\tt 0805.3033}}.

\bibitem{Eynard2008}
B.~Eynard and M.~Marino, ``{A holomorphic and background independent partition
  function for matrix models and topological strings},''
  \href{http://arxiv.org/abs/0810.4273}{{\tt 0810.4273}}.

\bibitem{Giddings1986}
S.~B. Giddings and E.~J. Martinec, ``{Conformal Geometry and String Field
  Theory},''
\href{http://dx.doi.org/10.1016/0550-3213(86)90108-2}{{\em Nucl. Phys.} {\bf
  B278} (1986)  91}.

\bibitem{Giddings1987}
S.~B. Giddings and S.~A. Wolpert, ``{A TRIANGULATION OF MODULI SPACE FROM LIGHT
  CONE STRING THEORY},''
\href{http://dx.doi.org/10.1007/BF01215219}{{\em Commun. Math. Phys.} {\bf 109}
  (1987)  177}.

\bibitem{Gaiotto2009}
D.~Gaiotto, ``{N=2 dualities},'' \href{http://arxiv.org/abs/0904.2715}{{\tt
  0904.2715}}.

\bibitem{Marshakov1991}
A.~Marshakov, A.~Mironov, and A.~Morozov, ``{Generalized matrix models as
  conformal field theories: Discrete case},''
\href{http://dx.doi.org/10.1016/0370-2693(91)90021-H}{{\em Phys. Lett.} {\bf
  B265} (1991)  99--107}.

\bibitem{Kostov2007}
I.~K. Kostov, ``Conformal field theory techniques in random matrix models,''
  \href{http://arxiv.org/abs/hep-th/9907060}{{\tt hep-th/9907060}}.

\bibitem{Zamolodchikov1996}
A.~B. Zamolodchikov and A.~B. Zamolodchikov, ``{Structure Constants and
  Conformal Bootstrap in Liouville Field Theory},'' {\em Nucl.Phys. B} {\bf
  477} (1996)  577--605, \href{http://arxiv.org/abs/hep-th/9506136}{{\tt
  hep-th/9506136}}.

\bibitem{Dorn1994}
H.~Dorn and H.~J. Otto, ``{Two and three-point functions in Liouville
  theory},'' {\em Nucl.Phys. B} {\bf 429} (1994)  375--388,
  \href{http://arxiv.org/abs/hep-th/9403141}{{\tt hep-th/9403141}}.

\bibitem{Teschner2001}
J.~Teschner, ``Liouville theory revisited,'' {\em Class.Quant.Grav.} {\bf 18}
  (2001)  R153--R222, \href{http://arxiv.org/abs/hep-th/0104158}{{\tt
  hep-th/0104158}}.

\bibitem{Pestun2007}
V.~Pestun, ``{Localization of gauge theory on a four-sphere and supersymmetric
  Wilson loops},'' \href{http://arxiv.org/abs/0712.2824}{{\tt 0712.2824}}.

\bibitem{Belavin1984}
A.~A. Belavin, A.~M. Polyakov, and A.~B. Zamolodchikov, ``{Infinite conformal
  symmetry in two-dimensional quantum field theory},''
\href{http://dx.doi.org/10.1016/0550-3213(84)90052-X}{{\em Nucl. Phys.} {\bf
  B241} (1984)  333--380}.

\bibitem{kaneko}
J.~Kaneko, ``{Selberg Integrals and Hypergeometric Functions Associated with
  Jack Polynomials},'' {\em Siam. J. Math. Anal.} {\bf 24, No. 4} (1993)
  1086--1110.

\end{thebibliography}\endgroup

\end{document}